\documentclass{article}
\providecommand{\keywords}[1]{\textbf{Keywords:} #1}

\usepackage{graphicx}                  
\usepackage{subcaption}                
\usepackage{amsmath, amssymb, bm}      
\usepackage{breqn}                     
\usepackage{float}                     
\usepackage{siunitx}                   
\usepackage{booktabs, multirow}        
\usepackage{tikz}                      
\usepackage{listings}                  
\usepackage{appendix}                  
\usepackage{eso-pic}                   
\usepackage{algorithm}                 
\usepackage{algpseudocode}             
\usepackage{hyperref}                  

\usepackage[
  backend=biber,
  style=numeric-comp,
  sorting=none
]{biblatex}
\DeclareUnicodeCharacter{0435}{e}
\DeclareUnicodeCharacter{0440}{r}
\DeclareUnicodeCharacter{0443}{u}
\DeclareUnicodeCharacter{0412}{V} 
\DeclareUnicodeCharacter{0441}{s} 
\DeclareUnicodeCharacter{0442}{t} 
\DeclareUnicodeCharacter{043D}{n} 
\DeclareUnicodeCharacter{0438}{i} 
\DeclareUnicodeCharacter{043A}{k} 
\DeclareUnicodeCharacter{0422}{T} 
\DeclareUnicodeCharacter{043E}{o} 
\DeclareUnicodeCharacter{043C}{m} 
\DeclareUnicodeCharacter{0433}{g} 
\DeclareUnicodeCharacter{0434}{d} 
\DeclareUnicodeCharacter{0430}{a} 
\DeclareUnicodeCharacter{0432}{v} 
\DeclareUnicodeCharacter{043F}{p} 
\DeclareUnicodeCharacter{0447}{ch} 

\begin{document}

\title{Loop-Corrected Scalar Potentials and Late-Time Acceleration in \(f(R)\) Gravity
}
\author{
  Pradosh Keshav MV\thanks{Email: pradosh.keshav@res.christuniversity.in} \\
  Kenath Arun\thanks{Email: kenath.arun@christuniversity.in} \\
  Department of Physics and Electronics \\ Christ University, Bangalore, India-560029
}

\maketitle

\begin{abstract}
We construct an analytic $f(R)$ gravity model that unifies early-time inflation with late-time cosmic acceleration within a single covariant framework. At high curvature, the model reproduces a Starobinsky-like inflationary plateau, while at low curvature it asymptotes to a stable dark energy–dominated phase. In the scalar-tensor representation, this construction yields a hilltop-type potential in the Jordan frame, which maps to an exponential potential in the Einstein frame. To account for radiative effects, we introduce a logarithmic correction to the Einstein-frame potential inspired by one-loop effective field theory, producing a late-time flattening without requiring fine-tuning. The resulting scalaron dynamics reduce the effective mass to $\mathcal{O}(H_0)$, inducing a thawing regime that deviates from a cosmological constant at the sub-percent levels. A joint background likelihood analysis using Pantheon+SH0ES and BAO+CC datasets (within the CPL parametrization) yields $H_0 = 73.4 \pm 0.6$ km/s/Mpc and $\Omega_m = 0.253 \pm 0.007$, consistent with local expansion rate measurements. The best-fit scalar field parameters are $\varphi_0 \approx 0.027\,M_{\rm Pl}$ and $\lambda \approx 0.010\,M_{\rm Pl}$, corresponding to a present-day dark energy equation of state $w_0 \approx -0.985$. While compatible with $\Lambda$CDM within current observational bounds, the model satisfies GR recovery at low curvature and exhibits attractor-like behavior, thereby minimizing sensitivity to initial conditions. 

\end{abstract}
\section{Introduction}

Cosmic inflation is considered to be the most compelling resolution to the flatness, horizon, and monopole problems of standard Big Bang cosmology. Simultaneously, it provides the origin of the nearly scale-invariant spectrum of primordial curvature perturbations that seeded the large-scale structure of the universe \cite{baumann2009cosmological, liddle2000cosmological}. This early-time quasi-de Sitter expansion is typically modeled by a single scalar field slowly rolling down a potential, and is supported by increasingly precise measurements of the Cosmic Microwave Background (CMB) temperature anisotropies and polarization \cite{aghanim2020planck}. The post-inflationary universe undergoes a reheating phase where the inflaton oscillates about the minimum of its potential, converting vacuum energy into relativistic particles and initiating the thermal history \cite{Albrecht1982, Abbott1982, ford1987gravitational, Kofman1994, Kofman1997, Shtanov1995}. However, recent observations have also confirmed that the universe has entered a second phase of accelerated expansion at late times \cite{riess1998observational, perlmutter1999measurements}, suggestive of a new component of energy density with strongly negative pressure—commonly referred to as dark energy. Since the cosmological constant $\Lambda$ remains the simplest candidate, its interpretation as vacuum energy is plagued by a severe fine-tuning problem—over 120 orders of magnitude between theory and observation—and by the coincidence that dark energy becomes dominant only at the current epoch \cite{arun2017dark}.

While scalar field models effectively describe both cosmic acceleration phases, their classical potentials often require fine-tuning to specific functional forms, such as runaway, hilltop, or exponential potentials, to account for both early and late-time dynamics \cite{Copeland1998}. This has led to extensive exploration of dynamical dark energy models, particularly those involving scalar fields, collectively known as quintessence \cite{zlatev1999quintessence}. These models offer an appealing alternative in which the present acceleration arises from the dynamics of a slowly evolving scalar field. Runaway potentials, in particular, are integral to quintessential inflation frameworks that rely on non-standard reheating mechanisms \cite{Felder1999, Lyth2002, delCampo2009, Hardwick2016, Dimopoulos2018}. Early studies, such as by Ford \cite{ford1987gravitational} and Spokoiny \cite{spokoiny1993deflationary}, demonstrated that gravitational particle production at the end of inflation could lead to entropy generation without conventional reheating. The idea of “quintessential inflation,” first proposed by Peebles and Vilenkin \cite{peebles1999quintessential}, is a particularly elegant realization of this concept, in which a single scalar field survives post-inflationary evolution to later become quintessence. These models necessarily forego standard reheating, as the scalar must remain undiluted, leading to alternative mechanisms such as gravitational particle production \cite{ford1987gravitational, spokoiny1993deflationary}, Ricci-induced matter creation, or instant preheating \cite{Felder1999, Lyth2002, delCampo2009}. The resulting kinetic-dominated epoch—kination—leads to distinctive observational signatures, such as enhanced relic gravitational waves \cite{Hardwick2016, Dimopoulos2018}.

Early tracker models with inverse power-law potentials were shown to exhibit attractor-like behavior, rendering the late-time evolution insensitive to initial conditions. These models belong to the broader class of freezing quintessence, in which the scalar field rolls rapidly at early times and gradually slows down, asymptotically approaching a cosmological constant–like behavior ($w_\phi \to -1$) \cite{zlatev1999quintessence, Steinhardt1999, pantazis2016comparison}. Freezing model \cite{dutta2011slow}, including those based on scaling solutions, has the advantage of tracking the dominant background component for a wide range of initial conditions. However, they require finely balanced potentials to ensure a late-time transition to acceleration, and often predict a stronger evolution of $w_\phi(z)$ than allowed by data. Indeed, high-precision cosmological datasets—including recent results from the Dark Energy Spectroscopic Instrument (DESI) \cite{lodha2025desi, gialamas2025interpreting}—increasingly disfavor strong tracking behavior and restrict the viable phase space to models where $w_\phi(z)$ remains close to $-1$. This has shifted attention to thawing quintessence scenarios \cite{caldwell2005limits, chiba2009slow}, in which the scalar field remains Hubble-frozen for most of cosmic history and only begins to evolve recently ($z \lesssim 2$). These models are theoretically well-motivated and naturally compatible with current observations, yet they introduce a new challenge: the field’s late-time dynamics are sensitive to its initial conditions, and departures from $w = -1$ are often too small to distinguish from $\Lambda$CDM \cite{tada2024quintessential, de2025thawing, ramadan2024desi}. Most thawing potentials—such as  PNGB, exponential plateau, or quadratic hilltop—predict $w_0 \gtrsim -0.98$, leading to suppressed signatures in the Hubble expansion rate and growth observables. Consequently, despite their theoretical appeal, thawing models have yet to deliver statistically significant improvements over the cosmological constant \cite{wolf2024scant}, unless non-trivial modifications to structure growth or earlier field evolution are introduced. 

Among these, $f(R)$ models have garnered particular attention due to their additional scalar degrees of freedom without explicit matter couplings, as exemplified by early proposals such as the Starobinsky inflationary scenario and its subsequent generalizations \cite{capozziello2006cosmological, capozziello2006dark, capozziello2006f, capozziello2011extended}. Several proposals have demonstrated that a suitably reconstructed $f(R)$ Lagrangian can mimic inflationary expansion at high curvature and drive late-time acceleration at low curvature, without introducing an explicit cosmological constant \cite{singh2003unified, Gannouji2012, Cosmai2016}.  Assuming a spatially flat FRW universe, the modified Friedmann equations in \( f(R) \)-gravity introduce an effective equation of state (EoS),
\begin{equation}
    w = -1 - \frac{2\dot{H}}{3H^2},
\end{equation}
where \( H \) denotes the Hubble parameter. In many models, the EoS parameter \( w \) needs careful tuning to match the conditions for both inflation and dark energy. For power-law dependencies \( f(R) \propto R^n \), approximate solutions demonstrate distinct inflationary and dark energy regimes, though fine-tuning is often necessary:
\begin{equation}
H(t) \sim
\begin{cases}
-\dfrac{(n-1)(2n-1)}{(n-2)\,t}, & \text{Phantom-like} \\ 
\;\;\dfrac{2n}{3(w+1)\,t},     & \text{GR with barotropic fluid}
\end{cases}
\label{eq:Hcases}
\end{equation} The first expression corresponds to a phantom-like regime in modified $f(R)$ gravity, where $H < 0$ indicates super-accelerated expansion \cite{nozari2009phantom}. The second is the standard General Relativity (GR) scaling solution for a barotropic fluid with equation of state $p = w\rho$ \cite{saez2011stability}. While $f(R)$ models can unify early- and late-time acceleration without invoking additional scalar fields—attributing inflation and dark energy to spacetime curvature—they typically require fine-tuning of parameters to achieve a stable late-time de Sitter phase \cite{appleby2010curing}. For instance, Starobinsky’s model \cite{starobinsky1980new} demonstrates that higher-order Ricci scalar corrections can generate inflation purely geometrically. However, ensuring a smooth transition from high-curvature (inflationary) to low-curvature (dark energy) regimes often necessitates delicate balancing of the functional form of $f(R)$. Moreover, these models must suppress deviations from Newtonian gravity to satisfy local tests, posing further challenges to their viability \cite{erickcek2006solar, dolgov2003can, chiba20031}.

One critical challenge in unifying Starobinsky inflation with quintessential inflation is that in the standard Starobinsky model, the scalaron (which emerges from the $R^2$ term) drives inflation but cannot naturally become quintessence, as it would conflict with local gravity tests. The scalaron's effective mass, $m \sim 10^{13}$ GeV, is orders of magnitude too large to account for the present-day dark energy scale, $m_{\text{DE}} \sim H_0 \sim 10^{-33}$ eV. Moreover, after inflation ends, the scalaron decays efficiently via coherent oscillations, reheating the universe through gravitational particle production. Retaining such a massive scalar degree of freedom at late times would generically lead to observable deviations from GR, in direct conflict with stringent local gravity constraints \cite{hoyle2004submillimeter}. An alternative approach was proposed in \cite{dimopoulos2021quintessential}, which attempted to realize a unified inflation–quintessence model in the Palatini formulation of gravity, avoiding the introduction of additional scalar fields. However, it remains unclear whether Palatini gravity, which lacks a dynamical scalaron in the Einstein frame, can support a successful inflationary plateau while accommodating the steep potentials required for quintessential behavior. The absence of a coherent scalar degree of freedom further limits the analysis of perturbations and the implementation of reheating, rendering such models incomplete or observationally inaccessible \cite{Opferkuch2019}.  In scenarios where the post-inflationary potential becomes too steep, the scalar field dilutes rapidly and cannot act as a viable quintessence candidate without invoking non-standard reheating mechanisms or additional tuning. Although one could tackle this by introducing a new reheating mechanism for the field to survive till today, the core issue remains: any realistic $f(R)$ embedding of this framework tends to make the scalaron unacceptably massive at late times, precluding it from sourcing dark energy while remaining consistent with both early-universe dynamics and late-time gravity tests.

In this paper, we propose a metric $f(R)$ gravity model with odd powers $(R-R_0)^{2n+1}$ instead of Starobinsky's $R^2$ term, which, however emerges from a geometric scalar potential \(V(\phi)\)  where $\phi \equiv f'(R)$. This avoids the problematic high-energy behavior of the quadratic  $R^2$ term that renders the scalaron excessively massive, precluding it from being dominant at late times. The odd-power construction gives rise to a high-curvature plateau suitable for inflation and a distinct low-curvature minimum that supports late-time acceleration, such that the scalar field survives to the present epoch without violating local gravity constraints. The emergent scalar degree of freedom \(\phi\) is introduced via $f'(R)$, similar to scalaron, but is endowed with a hilltop-type potential corrected by loop-level logarithmic terms \cite{ketov2015f}. These corrections are useful so that the field remains trapped near the minimum, avoiding the instability of rolling to zero, and preserving a shallow slope at low energies \cite{buchbinder2017effective}. As a result, the effective scalar mass remains sufficiently small in the infrared, enabling the field to act as quintessence. The model is reconstructed to dynamically recover general relativity in the low-curvature regime, specifically at \(R = R_0\), where the consistency conditions $f(R_0) \approx R_0$ and $f'(R_0) \approx 1$ are satisfied. Our primary aim is to provide a unified and analytically controlled description of both early-time inflation and late-time dark energy, without invoking fine-tuned matching or piecewise model construction. Finally, to assess the observational viability of the model, we confront it with current data from supernovae, cosmic chronometers, and baryon acoustic oscillations, and verify that its predictions remain compatible with \(\Lambda\)CDM-like expansion history within the current observational precision.

We discuss two fundamental challenges in constructing viable \(f(R)\) theories of gravity. The first is the stabilization of the scalar degree of freedom at late times while ensuring that the function \(f(R)\) satisfies the stability condition \(f''(R) > 0\) across all curvature scales. We resolve this by introducing a loop-corrected hilltop potential in the scalar-tensor frame, where the logarithmic term ensures a shallow slope in the infrared, avoiding tachyonic instabilities and preventing the field from rolling to zero. The second challenge lies in reheating: the absence of a true minimum in the runaway potential precludes standard reheating via scalar oscillations. While this work does not implement a specific reheating mechanism, it highlights the potential for Ricci reheating, where energy transfer occurs through a secondary scalar field \(\chi\) non-minimally coupled to the curvature. Such a mechanism can generate sufficient radiation while leaving the primary scalar field inert and available to source late-time acceleration. These results demonstrate that it is possible to construct a unified, analytic \(f(R)\) framework that interpolates between inflation and dark energy with quantum stability, while remaining compatible with current observational bounds. Further work will explore perturbation-level constraints and explicit reheating scenarios within this class of models.

The paper is structured as follows: In Section II, we present the scalar–tensor representation of $f(R)$ gravity and motivate the use of a hilltop potential derived from an odd-power curvature ansatz. Section III develops the core construction of the model, introducing loop corrections to the Einstein-frame potential and analyzing their implications for inflation and dark energy. Section IV outlines the reconstruction procedure for recovering the underlying $f(R)$ function and verifying its consistency. Section V focuses on the late-time dynamics, deriving the evolution of the scalar field, the effective equation of state, and the loop-corrected scalaron mass. In Section VI, we compare the model with observational data from Pantheon+SH0ES and BAO+CC datasets, providing constraints on key parameters such as $H_0$, $w_0$, and $\Omega_m$. We conclude in Section VII with a discussion of the model’s implications, observational viability, and prospects for further development.

\section{Early-time dynamics from \(f(R)\)-Gravity }
We begin by considering the Jordan-frame action for \(f(R)\)-gravity:
\begin{equation}
\label{eq:action}
S = -\frac{1}{2\kappa} \int d^4 x \sqrt{-g} \left[ \phi R - V(\phi) \right], \quad \kappa^2 = {8\pi G},
\end{equation}
which introduces a scalar degree of freedom \(\phi \equiv f'(R)\) dynamically equivalent to \(f(R)\)-gravity \cite{ferraro2012f}. Variation with respect to \(\phi\) yields the constraint equation \(R = V'(\phi)\), tethering spacetime curvature to the potential's gradient where \(V(\phi)\) is postulated a priori. By introducing \(\phi\equiv f'(R)\) and performing the Legendre transform, we obtain:
\begin{equation}
f(R) = \phi R - V(\phi),\label{eq:req2}
\end{equation}where \(f(R)\) immediately exhibits the scalar potential \(V(\phi)\) in the Jordan frame (see Appendix A). During slow roll, the friction term \( 3H\dot{\phi}\) necessitates violating the convexity condition \(V''(\phi) > 0\) \cite{felder2002cosmology}. From the averaged stress-energy components, acceleration requires:
\begin{equation}
\label{eq:nonconvex_condition}
\phi V'(\phi) - V(\phi) > 0.
\end{equation}
This condition must hold during inflation, i.e, the potential must be tailored to satisfy this in the inflationary region demanding non‑convexity near \(\phi\sim\phi_c\). Thus, we need potentials which are sufficiently non-convex near the maximum amplitude of \(\phi\) to overcome the ‘drag’ from the potential slope. Then inflation continues for \(0<\phi<\phi_c\), and one has \(\ddot a>0\). Hence, the non‐convex (hilltop) region that sustains inflation is \(\phi\) inside the interval \((0<\phi<\phi_c)\).

\begin{figure}[ht!]
    \centering
    \begin{subfigure}[t]{0.45\textwidth}
        \includegraphics[width=\textwidth]{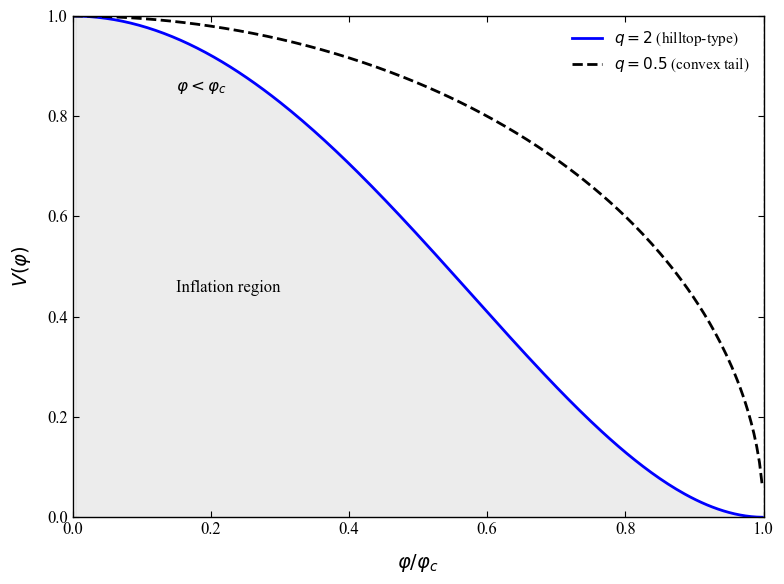}
        \caption{\small Hilltop potential \(V(\phi)\) for \(q = 2\) (solid) vs. convex quintessence potential \(q = 0.5\) (dashed). The shaded region \(\phi < \phi_c\) satisfies Eq. \eqref{eq:nonconvex_condition}, enabling sustained inflation.}
        \label{fig:potential}
    \end{subfigure}
    \hfill
    \begin{subfigure}[t]{0.45\textwidth}
        \includegraphics[width=\textwidth]{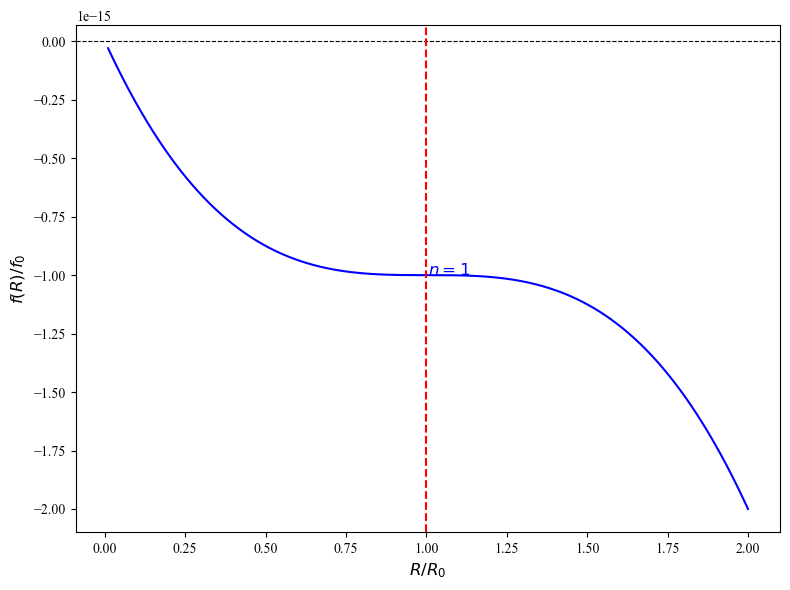}
        \caption{\small \(n=1\): Gradual transition to \(\Lambda_i\).}
        \label{fig:n1}
    \end{subfigure}
    \hfill
    \begin{subfigure}[t]{0.45\textwidth}
        \includegraphics[width=\textwidth]{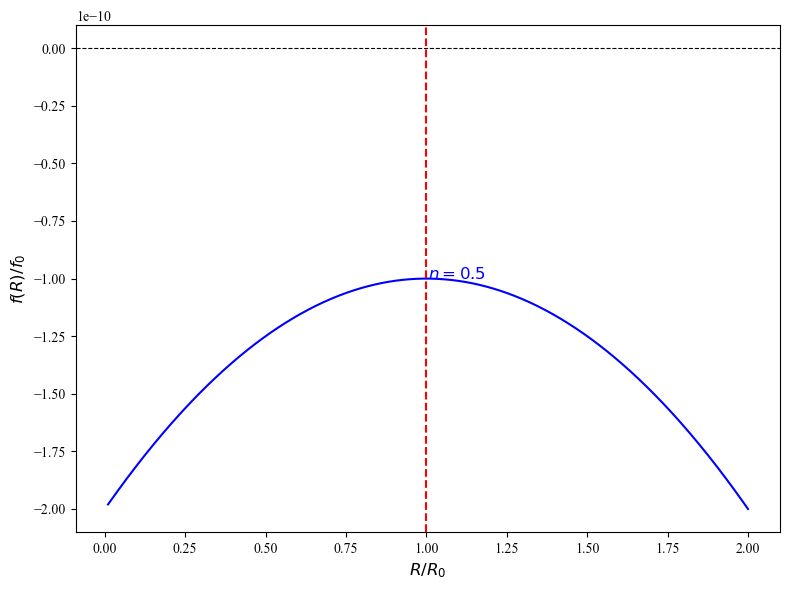}
        \caption{\small \(n=2\): Steeper inflationary plateau, suppressing \(R\)-oscillations.}
        \label{fig:n2}
    \end{subfigure}
    \caption{\small (a) Acceleration \(\ddot{a}>0\) holds for \(0<\phi<\phi_c\) (shaded region). The potential for \(q=2\) features a convex minimum at \(\phi_c\) to facilitate reheating. (b,c) \(f(R)/f_0\) vs. \(R/R_0\) for varying \(n\). The model is regular at $R = R_0$ and flattens at low curvature, mimicking a cosmological constant. The red dashed line marks the reference scale $R = R_0$. For viable late-time acceleration, the modification must satisfy $f_{RR} > 0$ and be consistent with the scalaron mass bound \cite{odintsov2025dynamical}.}
    \label{fig:fR_behavior}
\end{figure}

To explicitly satisfy the latter condition and maintain a quasi-flat inflationary plateau, we adopt a class of hilltop potentials where the curvature \(V''(\phi)\) becomes negative in the inflationary regime (see \cite{boubekeur2005hilltop, lillepalu2023generalized}). For substituting \(R = V'(\phi)\) in Eq.\eqref{eq:req2}, we adopt a quadratic $(m=2)$ hilltop form:
\begin{equation}
  V(\phi) = A \left[\left(\frac{\phi}{\phi_c}\right)^{2} - 1\right]^q, \label{eq:quintpot}
\end{equation}
\begin{equation}
     V'(\phi) = {2Aq} \left( \frac{\phi}{\phi_c^{2}} \right) \left[ \left( \frac{\phi}{\phi_c} \right)^2 - 1 \right]^{q-1}, \label{eq:potentialderivative}
\end{equation}where \(A\) is the effective amplitude of the potential given by the shape parameters: $\phi_c$ is the characteristic field scale, and $q > 1$ determines the flatness near the hilltop. In order to ensure that, the first and second derivatives of the potential in Eq. \eqref{eq:quintpot} need to be constrained, we require \(V'(0)_{\rm HI} = 0\) and \(V''(0)_{\rm HI} < 0\). Figure \ref{fig:potential} contrasts this with standard quintessence potentials, demonstrating how \(q > 1\) flattens \(V(\phi)\) near \(\phi_c\), satisfying Eq.\eqref{eq:nonconvex_condition} for \(\phi<\phi_c\). In the current form, the potential exhibits a plateau-like structure where the potential slow-roll parameters:
\begin{equation}
\epsilon_V(\phi) = \frac{M_{\rm Pl}^2}{2} \left( \frac{V'}{V} \right)^2,
\qquad 
\eta_V(\phi) = M_{\rm Pl}^2 \frac{V''}{V}.
\end{equation}interpolates between large-field inflation (power-law behavior) for \(\phi \gg \phi_c\) and hilltop-like inflation near the maximum at \(\phi \sim \phi_c\). The behavior of the slow-roll parameters—and thus observables such as $n_s$ and $r$—depends sensitively on the field value $\phi_*$ at which inflation begins. For small $\phi_*$, the potential is steep and curved, while for $\phi_* \gg \phi_c$, it flattens asymptotically. This is qualitatively similar to Starobinsky inflation \cite{ivanov2022analytic} or \(\alpha\)-attractor models \cite{galante2015unity, garcia2018dark}, where the potential is flat at large \(\phi\) and observables saturate at universal values for large fields. For instance, towards the end of inflation $\epsilon_V(\phi_{\rm end}) = 1$, one could solve numerically to obtain $\phi_{\rm end} \approx 1.93\,M_{\rm Pl}$ for $\phi_c = M_{\rm Pl}$. The total number of e-foldings from a given initial field value $\phi_*$ is
\begin{align}
N(\phi_*) &= \frac{1}{M_{\rm Pl}^2} \int_{\phi_{\rm end}}^{\phi_*} \frac{V}{V'}\, d\phi \nonumber \\
&= \frac{1}{4M_{\rm Pl}^2} \left( \phi_*^2 - \phi_{\rm end}^2 \right) - \frac{\phi_c^2}{2M_{\rm Pl}^2} \ln\left( \frac{\phi_*}{\phi_{\rm end}} \right).
\end{align}For moderate field values ($\phi_* \sim 15\,M_{\rm Pl}$), the model yields a spectral index $n_s \sim 0.964$ consistent with Planck data, but predicts a relatively large tensor-to-scalar ratio $r \sim 0.14$, which exceeds the current upper bound $r < 0.06$. In contrast, pushing the initial field value deeper into the flat tail of the potential ($\phi_* \gtrsim 25\,M_{\rm Pl}$) significantly suppresses the tensor amplitude ($r \sim 0.05$), satisfying observational limits, but at the cost of producing a scalar spectral index $n_s \sim 0.987$, which is higher than allowed by CMB measurements.

This trade-off originates from the asymptotically flat structure of the hilltop potential, which suppresses both $\epsilon_V$ and $\eta_V$ as $\phi_*$ increases, driving the spectral tilt $n_s \to 1$. While such flattening is effective at reducing the tensor-to-scalar ratio $r$, it also leads to a nearly scale-invariant spectrum that exceeds current observational bounds. Thus, the model imposes a tension between suppressing tensor modes and maintaining a realistic scalar tilt, constraining the allowed field excursion during inflation. To reconcile this, one can consider adjusting the power-law index $q$ to increase curvature near the plateau edge, or introducing additional terms (e.g., exponential flattening or loop corrections) that naturally truncate the tail behavior. These refinements preserve the geometric origin of the potential while enabling compatibility with Planck constraints. In this paper, we will focus on one-loop corrections, which are particularly appealing in the broader context of unifying early- and late-time acceleration through a single $f(R)$-derived framework. We will discuss this in detail in the upcoming sections. 

Next, we will construct an $f(R)$ ansatz designed to mimic the desired inflationary flattening while introducing a dynamical mechanism to regulate the transition between high- and low-curvature regimes. The functional form of $f(R)$ model is expected to provide analytic control over the steepness of the inflationary plateau (as discussed previously) via the odd-power term $(R - R_0)^{2n+1}$. This approach will ensure regularity at \(R = R_0\) while facilitating a smooth transition to a low-curvature, late-time accelerating universe. The corresponding \(f(R)\) ansatz takes the form:
\begin{align}
\label{eq:fR_ansatz}
f(R) &= -\frac{
(R - R_0)^{2n+1} + R_0^{2n+1}
}{
f_0 + f_1 \left[(R - R_0)^{2n+1} + R_0^{2n+1}\right]
}, \quad f_0 > 0.
\end{align}
The above equation includes Starobinsky’s odd-power terms, thereby avoiding branch-cut singularities often found in even-power models \cite{starobinsky1980new}. This allows for a built-in deformation of the asymptotic tail that suppresses the overproduction of scale-invariant perturbations—an issue inherent in many minimally modified plateau models—and replaces fine-tuned potential design with geometric structure \cite{hoffmann2021squared}. 

Asymptotically, \(f(R) \to -\Lambda_i \equiv -1/f_1\) for \(R \to \infty\) (inflationary de Sitter phase) and \(f(R) \to -2\tilde{R}_0\) near \(R \sim R_0\) (late-time acceleration), with \(\tilde{R}_0 \sim H_0^2\). Parameters are fixed by Planck constraints \cite{aghanim2020planck}:
\begin{align}
    f_1 &\sim \Lambda_i^{-1}, \quad \Lambda_i \sim (10^{15}\ \text{GeV})^4, \nonumber \\
    f_0 &= R_0^{2n+1} \left(\frac{1}{2\tilde{R}_0} - f_1\right), \quad R_0 = 12H_0^2.
\end{align}
where \(\tilde{R}_0\) is determined from late-time observations. The steepness parameter \(n \geq 1\) suppresses high-curvature deviations from \(\Lambda_i\), as shown in Fig. \ref{fig:fR_behavior}, while larger \(n\) steepens the transition to the inflationary plateau, aligning with CMB bounds on scalar perturbations.  Although \(f(R)\) driven inflation at high curvature asymptote to a constant, small oscillations in \(R\) result in small variations in \(f(R)\), which can still sustain the same hilltop form as long as these oscillations are around a central value that sustains a quasi-de Sitter expansion. 

Having established the structure of the $f(R)$ function, we now turn to the dynamical evolution of the scalar field $\phi \equiv f'(R)$. Once the functional form of $f(R)$ is specified, the scalaron field $\phi$ acquires a well-defined potential $V(\phi)$ via Legendre transformation, and its evolution is governed by the Friedmann equations coupled to a modified Klein-Gordon equation. The dynamics of $\phi$ not only determine the inflationary expansion but also mediate the reheating phase that follows. In particular, the scalar field evolves according to:
\begin{equation}
\label{eq:KleinGordon}
\ddot{\phi} + 3H \dot{\phi} = -\frac{1}{3} \left( 2V(\phi) - \phi V'(\phi) \right).
\end{equation}
As \(R \to R_0\), the field \(\phi\) approaches a constant, and the universe transitions from inflation to the next phase. The scalar field \(\phi\) oscillates around \(\phi_c\), converting its energy into radiation—a reheating mechanism intrinsic to \(f(R)\)-gravity \cite{mathew2020primordial, felder2002cosmology}, bypassing ad-hoc couplings. At high curvature (\( R \gg R_0 \)), the function behaves effectively as a plateau-like term, supporting inflation in analogy with Starobinsky's \( R^2 \) gravity. Conversely, near \( R \sim 0 \), the nonlinearity softens and induces a residual vacuum curvature, mimicking dark energy without invoking a true cosmological constant. While Eq.~(\ref{eq:fR_ansatz}) is not analytically transformed into the loop-corrected potential used in the next sections, it provides the geometric origin for the scalar degree of freedom.

In this way, one could resolve two fine-tuning issues: (i) quintessence models' reliance on ad-hoc potentials is replaced by \(V(\phi)\)'s geometric origin from Eq. \eqref{eq:fR_ansatz}, and (ii) the denominator in Eq. \eqref{eq:fR_ansatz} guarantees \(f''(R) > 0\) universally, evading Dolgov-Kawasaki instabilities \cite{dolgov2003can} (leading to a negative effective mass squared for the scalaron) to avoid catastrophic gravitational runaway.  
\section{The Model}
In this section, we reframe Eq.\eqref{eq:fR_ansatz} as a suitable class of models for dual phase acceleration without a true cosmological constant but nonetheless include the phenomenology of the standard model as a limiting case. We perform the conformal transformation to the Einstein frame, defining: 
\begin{equation}
    \varphi = \sqrt{\tfrac32}\,M_{\rm Pl}\,\ln\phi,
\quad M_{\rm Pl}^2=(8\pi G)^{-1}\,,
\end{equation} and  
\begin{equation}
    \bar g_{\mu\nu}=\phi\,g_{\mu\nu},
\quad
\sqrt{-g}=\phi^{-2}\,\sqrt{-\bar g}\,.
\end{equation}Recall that \(\phi\) is the Jordan‑frame field (from Sec.2) and we now define the Einstein‑frame scalar via: \begin{equation}
    \phi(\varphi) = \exp\!\left(\sqrt{\tfrac{2}{3}}\,\frac{\varphi}{M_{\rm Pl}}\right) \label{eq:exponentialform}
\end{equation} so that all remaining formulae are expressed solely in terms of $\varphi$.  For instance, the hilltop scale $\phi_c$ will be replaced later in the section by $\varphi_c$, and every occurrence of $\phi$ in $V(\phi)$ or $f'(R)$ shall be understood as $e^{\sqrt{2/3}\,\kappa\varphi}$ where \(\kappa = M_{\rm pl}^{-1}\). 

Recent efforts by Odintsov et al. \cite{odintsov2025dynamical} have analyzed the behavior of the effective scalaron mass in the Einstein frame, which plays a crucial role in determining the stability of de Sitter perturbations. This mass is computed using the formula
\begin{equation}
    m_\varphi^2(R) = \frac{1}{3} \left( \frac{f'(R)}{f''(R)} - R \right) \label{eq:massform}
\end{equation} which corresponds to the squared mass of the scalar degree of freedom arising in metric $f(R)$ gravity. This is equivalent to the canonical scalaron mass $m_\phi^2 \equiv d^2U/d\phi^2$, where $U(\phi)$ is the potential defined in the Einstein frame. However, to avoid tachyonic instabilities, a model-independent condition must be satisfied: 
\begin{equation}
    0 \leq \frac{Rf''(R)}{f'(R)} \leq 1 \label{eq:16}
\end{equation}which constrains the allowed form of $f(R)$ near de Sitter fixed points. This condition applies to any \(f(R)\) theory that aims to unify early and late-time cosmic acceleration within a single scalar-tensor framework \cite{odintsov2023recent}.

Taking this context into consideration, the Jordan‑frame curvature term transforms as  \begin{equation}
    R
=\phi^{-1}\Bigl[\bar R
 -3\,\bar\square\ln\phi
 -\tfrac32\,\bar g^{\mu\nu}\partial_\mu\ln\phi\,\partial_\nu\ln\phi
\Bigr],
\end{equation}
so that the action in the Einstein frame becomes  
\begin{equation}
    S' 
=\frac{1}{2\kappa^2}\!\int\!d^4x\,\sqrt{-\bar g}
\Bigl[\bar R - (\bar\nabla\varphi)^2 -2\,\kappa^2\,U(\varphi)\Bigr],
\end{equation}
with
\begin{equation}
    U(\varphi)
=\frac{V(\phi(\varphi))}{2\kappa^2\,\phi(\varphi)^2} 
\, \label{eq:15}
\end{equation}which will be rendered explicit once \(U(\varphi)\) is obtained. Here, \(V(\phi)\) arises from the Legendre transform of the original \(f(R)\) action and the potential curvature $U''(\varphi)$ directly controls the scalaron mass. A positive and slowly varying $U''(\varphi)$ implies that $m^2_\varphi > 0$, where Eq. \eqref{eq:16} will be satisfied.

To implement this designer ansatz using Eq.\eqref{eq:fR_ansatz}, we take \(R=V'(\phi)\) from Section 2.  Setting: 
\begin{align}
    X &\equiv R - R_0, \nonumber \\
    g(X) &= X^{2n+1} + R_0^{2n+1}, \nonumber \\
    g'(X) &= (2n+1)\, X^{2n}
\end{align}
one has  
\begin{align}
    f(R) &= -\frac{g(X)}{f_0 + f_1\, g(X)}, \nonumber \\
    f'(R) &= \frac{g'(X)\, (f_0 + f_1\, g) - g\, f_1\, g'(X)}{(f_0 + f_1\, g)^2} \nonumber \\
          &= - (2n + 1)\, X^{2n} \, \frac{f_0}{\left[f_0 + f_1\, g(X)\right]^2}.
\end{align}
A non‑zero \(f'(R)\) parameterizes deviations from the Einstein–Hilbert term; for large \(\phi\) (hence large \(R\)) the odd‑power structure yileds a smooth interpolation between the inflationary plateau \(-1/f_1\) and the late‑time value \(-2\tilde R_0\). Interestingly, this structure allows exact numerical control over the Odintsov variable \(y(R) = R f''(R)/f'(R)\) (see \cite{odintsov2025dynamical}, \cite{oikonomou2021power}) where one can tune the parameters \(n, f_0,\) and \(f_1\) such that $y(R \to R_{\text{inf}}) \approx 0$ (where the scalaron becomes heavy during inflation) and $y(R \to R_0) \approx 1$ (where the scalaron mass \(m_\varphi^2 \to 0\), recovering slow-roll dark energy).

For the special case \(q=2\) in Eq. \eqref{eq:potentialderivative}, we use \(R=V'(\phi)\) to obtain:  
\begin{dmath}
    f'(R(\phi)) = - (2n+1)\, \left( \frac{4A}{\phi_c^4} \phi (\phi^2 - \phi_c^2) - R_0 \right)^{2n} \cdot
\frac{f_0}{\left[ f_0 + f_1 \left( \left( \frac{4A}{\phi_c^4} \phi (\phi^2 - \phi_c^2) - R_0 \right)^{2n+1} + R_0^{2n+1} \right) \right]^2}.\label{eq:firstderi}
\end{dmath}One checks algebraically that the two terms in Eq. \eqref{eq:firstderi} combine to reproduce the compact form given in Section II. In the large-field limit $\phi \gg \phi_c$, we find $f'(R) \sim \phi^{-6} \to 0$, implying that the scalar degree of freedom becomes strongly coupled. However, this does not correspond to standard GR recovery, which would require $f'(R) \to 1$ and $f''(R) \to 0$. Rather, the vanishing of $f'(R)$ indicates deviation from GR and must be interpreted in the Einstein frame, where the field dynamics can remain perturbative. For $q = 2$, the potential in Eq. \eqref{eq:quintpot} exhibits a quadratic minimum near $\phi = \phi_c$, consistent with slow-roll dynamics in inflation, while larger values of $q$ yield steeper potentials and delayed thawing behavior at late times.

To maintain theoretical consistency with slow-roll behavior, we prefer the odd-power index $n = 1$ over even values. The case $n=1$ yields a smooth, single-hump hilltop profile for $f'(R(\phi))$, avoids singularities or cusps, and supports a well-behaved Einstein-frame potential with stable curvature corrections. In contrast, $n=2$ introduces excessive steepness in $f'(R)$ that destabilizes the field near the transition region. Moreover, odd powers ensure regularity around $R = R_0$ and align with the expected analytic structure of loop-corrected $f(R)$ expansions. The behavior of $f'(R)$ for $n = 0,1,2$ is systematically analyzed in Table \ref{tab:fr_behavior}, which shows that $n = 1$ provides an optimal balance between a viable inflationary mechanism and thawing dark energy dynamics at late times. A detailed derivation of the GR recovery and asymptotic behaviour at \(n=1\) is given in Appendix A.

\begin{table*}
\scriptsize
\centering
\begin{tabular}{lccc}
  \toprule
  \textbf{n} 
  & \textbf{Structure of \(f'(R(\phi))\)} 
  & \textbf{Peak Behavior} 
  & \textbf{Cosmological Implication} \\
  \midrule
  0 
  & Monotonic decay 
  & No peak 
  & Weak thawing, near \(\Lambda\)CDM \\
  1 
  & Smooth hilltop 
  & Broad peak at finite \(\phi\) 
  & Consistent with slow-roll dark energy \\
  2 
  & Sharp hilltop 
  & Narrow peak, faster descent 
  & Rapid thawing, potentially unstable \\
  \bottomrule
\end{tabular}
\caption{\small Behavior of \(f'(R(\phi))\) and resulting scalar field dynamics for different values \(n\) in the ansatz Eq. \eqref{eq:fR_ansatz}. The structure of \(f'(R(\phi))\) governs the shape of the effective potential \(U(\varphi)\) and the evolution of the scalar degree of freedom. Larger values of \(n\) induce steeper curvature near the origin and accelerate the field thawing from its frozen initial state.}
\label{tab:fr_behavior}
\end{table*}

The late-time behavior of the scalar field can include brief oscillatory phases, especially near the minimum of the potential. The time-averaged equation of state for a monomial potential $V(\phi) \propto \phi^{2q}$ is given by $\langle w \rangle = (q - 1)/(q + 1)$, implying that for $q > 1$, accelerated expansion can persist even during oscillatory regimes \cite{mukhanov2005physical}. This result supports the structural viability of the hilltop potential with $q > 1$ in maintaining negative pressure beyond the frozen phase. In particular, the cases $n = 1$ and $n = 2$ correspond to $q = 2$ and $q = 3$, yielding $\langle w \rangle = 1/3$ and $1/2$ respectively in the oscillatory regime, consistent with the qualitative trends seen in the steepness of the potential and the corresponding field dynamics. For instance, models with \(m=1\) correspond to standard hilltop inflation (see \cite{boubekeur2005hilltop, kallosh2019hilltop, lin2020topological, kohri2007more, german2021quartic, hoffmann2021squared} and refs. therein), described by the potential in Eq. \eqref{eq:quintpot}, while the configuration with \(m=2\) and \(q=2\) is equivalent to the symmetry-breaking model (see \cite{Olive:1989nu, kanti1999realization}). Since we are looking for reasonable restrictions on our model, we limit our analysis to \(0< q \leq 2\), as larger exponents lead to excessively steep potentials and fast oscillations that destabilize late-time evolution.

Substituting Eq. \eqref{eq:quintpot} into Eq. \eqref{eq:15} and using Eq. \eqref{eq:exponentialform} gives the Einstein frame equivalent for \(q=2\):  
\begin{equation}
U(\varphi) 
= \frac{A}{2\kappa^2}\,\biggl(
\frac{e^{2\sqrt{{2}/{3}}\kappa\varphi}}{\varphi_c^4}
\;-\;\frac{2}{\varphi_c^2}
\;+\;e^{-2\sqrt{{2}/{3}}\kappa\varphi}
\biggr)\,,\label{eq:upotenfinal}
\end{equation}
where one obtains an explicit quasi‑quadratic plus exponential tail potential in \(\varphi\). For \(\varphi\gg1\) (in reduced‑Planck units), the last term is negligible and we get: 
\begin{equation}
  U(\varphi) \simeq \frac{A}{2\kappa^2} \left( \frac{e^{2\sqrt{{2}/{3}}\kappa\varphi}}{\varphi_c^4} - \frac{2}{\varphi_c^2} \right). \label{eq:finaldepoten}
\end{equation}

Near $\varphi \approx \varphi_c$, the full potential $U(\varphi)$ in Eq.\eqref{eq:upotenfinal} admits a convex minimum suitable for reheating \cite{damour1998inflation}. Non‑linearities in \(U(\varphi)\) can, in principle, induce tachyonic directions or rapid oscillations since the contributing terms in Eq. \eqref{eq:upotenfinal} are proportional to the squared effective mass of the inflaton field, which is mostly negative. This tachyonic behavior is acceptable during inflation (when controlled by Hubble friction) but must be regulated post-reheating to avoid spoiling late-time cosmology. In the absence of slow‑roll, these must be controlled, for example, by introducing logarithmic flattening at late times, so that the field settles into a new effective minimum, preserving a stable dark‑energy epoch without unphysical oscillatory behavior.

While the Einstein-frame potential $U(\varphi)$ could reproduce the inflationary plateau and supports a post-inflationary minimum for reheating, its curvature near the origin may still induce instabilities or unphysical oscillations during late-time evolution. In particular, as the scalar field redshifts toward smaller values during cosmic expansion, the second derivative $U''(\varphi)$ can become negative in certain regions, leading to tachyonic growth or runaway solutions. To regulate this behavior and ensure a stable quasi-de Sitter vacuum at late times, we include a logarithmic correction term motivated by radiative loop effects. These one-loop contributions—analogous to the Coleman–Weinberg potential \cite{coleman1973radiative}—effectively modify the shape of $U(\varphi)$ in the small-field regime while preserving the inflationary plateau for large $\varphi$. We therefore consider the loop-corrected potential (to leading order) of the following form:

\begin{align}
\label{eq:loopcorrpoten}
U_{\rm corr}(\varphi) 
&= U(\varphi) + \mathcal{Z} \ln\left(\frac{\varphi}{\varphi_0}\right) \notag \\
&= \frac{A}{2\kappa^2\,\varphi_c^4} e^{2\sqrt{{2}/{3}}\,\kappa\varphi}
   - \frac{A}{\kappa^2\,\varphi_c^2}
   + \mathcal{Z} \ln\left(\frac{\varphi}{\varphi_0}\right)\,.
\end{align}
In this expression, the logarithmic term \(\mathcal{Z} \ln(\varphi/\varphi_0)\) mimics one-loop quantum corrections and is subdominant compared to the leading exponential terms during inflation and is therefore optional from a purely inflationary perspective. However, it plays a critical role (which we will show in upcoming sections) in regulating the curvature $U''(\varphi)$ near the origin while maintaining stability during the transition to the late-time quasi-de Sitter phase (see \cite{ballesteros2016radiative} and refs. therein). The constant term \( A/(2\kappa^2 \varphi_c^2) \) corresponds to a vacuum shift and does not influence the field dynamics at large \( \varphi \), where the exponential and logarithmic terms dominate. In practice, this term can be absorbed into the overall normalization of the potential or dropped in effective parametrizations where only the running behavior and scaling are physically relevant. The motivation for the loop-corrected potential, along with its effective field-theoretic origin in Eq. \eqref{eq:upotenfinal}, is shown in Appendix B.

To avoid singularities at $\varphi = 0$, we set $\varphi_0 > \varphi_c$, since the field remains confined to the hilltop region $0 < \varphi < \varphi_c$ during inflation and reheating. This means that the logarithmic correction remains subdominant and well-behaved throughout the relevant field range. Here, \( \mathcal{Z} \) is a parameter that controls the scale of the logarithmic correction, and \( \varphi_0 \) is a reference value for the scalar field. For $\mathcal{Z} = 0$, the potential reduces to its purely exponential form as given in Eq. \eqref{eq:finaldepoten}. However, for $\mathcal{Z} \neq 0$, the logarithmic term introduces a mild distortion to the potential, which can affect the curvature $U''(\varphi)$ i.e, a positive $\mathcal{Z}$ leads to a tachyonic instability in the limit $\varphi \ll \varphi_0$, as the second derivative becomes negative. In contrast, we choose $\mathcal{Z} < 0$, such that $U''(\varphi) > 0$ for all $\varphi > 0$. Provided $|\mathcal{Z}| \ll A$, the correction remains perturbative and does not significantly affect the slow-roll plateau, yet regulates the potential curvature in the small-field regime. This stabilizing logarithmic term also helps prevent the field from rapidly evolving to zero, supporting a sustained quasi-de Sitter phase.

\section{Recovering $f(R)$ gravity}
It has been shown in \cite{nojiri2006modified,nojiri2008dark} that with two auxiliary functions, one can reconstruct any desired cosmology in scalar–tensor form. We therefore merge \(U_{\rm corr}\) into the Jordan‑frame potential function and write  
\begin{equation}
\label{eq:S_Jordan}
S=\frac{1}{2\kappa^2}\int d^4x\sqrt{-g}\,\Bigl[P(\phi)\,R+\widetilde Q(\phi)+\mathcal L_{\rm matter}\Bigr].
\end{equation} The function \(\widetilde{Q}(\phi)\) is constructed from the corrected potential \(U_{corr}(\varphi(\phi))\), so that the Einstein–Jordan frame equivalence remains manifest. In the general scalar-tensor reconstruction, one writes \(P(\phi) \,R + Q(\phi)\) as the Jordan‑frame potential where we define  \(\widetilde Q(\phi)\;\equiv\;Q(\phi)\;+\;2\kappa^2\,U_{\rm corr}(\varphi)\) and absorbs the loop correction into the scalar–tensor potential. Note that throughout the reconstruction we will occasionally work in Planck units \( \kappa^2= 8 \pi G=1\), as the exact value might be absorbed into other parameters. 

We highlight a special case corresponding to the odd-power ansatz discussed in Sec .~III. In particular, we use \(n=1\), which leads to the Jordan-frame identification:
\begin{align}
    P(\phi) &= -\frac{f_0}{\left[ f_0 + f_1\,g(X) \right]^2} \cdot g'(X), \label{eq:Pphi_def} \\
    X &\equiv V'(\phi) - R_0, \label{eq:X_def} \\
    g(X) &= X^3 + R_0^3, \label{eq:gX_def}
\end{align}
where \( g'(X) = 3X^2 \). Variation with respect to \(\phi\) yields the algebraic “reconstruction” condition  
\begin{equation}
\label{eq:scalar_constraint}
0=P'(\phi)\,R+\widetilde Q'(\phi).
\end{equation}  The above equation \eqref{eq:scalar_constraint} algebraically fixes \(\phi=\phi(R)\).  Substituting back into the action yields the pure-\(f(R)\) form
\begin{equation}
\label{eq:fR_recon}
f(R)=P\bigl(\phi(R)\bigr)\,R+\widetilde Q\bigl(\phi(R)\bigr)\,.
\end{equation} Since \(\phi\) enters without a kinetic term in Eq. \eqref{eq:loopcorrpoten}, it is purely algebraic, and the back‑substituted combination in Eq. \eqref{eq:S_Jordan} gives the effective \(f(R)\) in the Jordan frame, such that it varies with respect to \(g_{\mu\nu}\) and gives the scalar-tensor field equation  
\begin{equation}
\label{eq:metric_variation}
P(\phi)\,G_{\mu\nu}
+\bigl[g_{\mu\nu}\,\Box-\nabla_\mu\nabla_\nu\bigr]P(\phi)
-\tfrac12\,\widetilde Q(\phi)\,g_{\mu\nu}
=\kappa^2\,T_{\mu\nu}.
\end{equation}  
Projecting Eq.~(\ref{eq:metric_variation}) onto the temporal component of the spatially flat FRW metric, one uses  
\(\;G_{00}=3H^2,\;G_{ij}=-(2\dot H+3H^2)a^2\delta_{ij},\;\Box P=\ddot P+3H\dot P\),  and reads off the two Friedman‑type equations:  
\begin{align}
\label{eq:Friedmann1}
3H^2\,P \;+\;3H\,\dot P\;+\;\frac12\,\widetilde Q 
\;=\;\kappa^2\,\rho,\\
\label{eq:Friedmann2}
-\bigl(2\dot H+3H^2\bigr)P \;-\;\ddot P\;-\;3H\,\dot P
\;-\;\frac12\,\widetilde Q
\;=\;\kappa^2\,p.
\end{align}  
Adding Eq.\,\eqref{eq:Friedmann2} with Eq.\,\eqref{eq:Friedmann1} eliminates \(\widetilde Q\). Then using the continuity equation \(\dot\rho+3H(\rho+p)=0\) yields the master reconstruction equation for \(F(\phi)\equiv P(\phi)\):  
\begin{equation}
\label{eq:master_raw}
 2 \,\ddot F \;-\;2\,H\,\dot F+\;4\,\dot H\,F
       + (\rho+p)
       \;=\;0,
\end{equation} where we have set the units \(\kappa^2=1\).

By imposing the gauge choice \(\phi=t\) (which restricts to non-phantom fields, i.e., \(\dot{\phi} > 0\) and allows cosmic-time parametrization) and using the ansatz \(\;a(t)=a_0e^{g(t)}\), we find that:
\begin{equation}
    H(\phi)=g'(\phi), \quad \text{and} \, \, \, \, \, \rho_i=\rho_{i0}a_0^{-3(1+w_i)}e^{-3(1+w_i)g(\phi)}
\end{equation} leading to a specific form: 
\begin{dmath}
\label{eq:master_recon}
2F''-2g'F'+4g''F+\sum_i(1+w_i)\rho_{i0}a_0^{-3(1+w_i)}e^{-3(1+w_i)g}=0,
\end{dmath}  
to which all subsequent reconstruction ansätze refer. Generalizations (such as \(\phi = \ln a\)) are reserved for future study. However, for any given cosmology, one can retain the specific \(f(R)\)-gravity model given that the reconstruction preserves the stability condition \(f''(R) > 0\). To show the master reconstruction ansätze, one could adopt different forms for \( g'(\phi) \) (see \cite{boisseau2000reconstruction, saez2009modified, odintsov2018reconstruction} for similar reconstruction methods).

Having reconstructed \( f(R) \) purely in terms of the Jordan-frame field \( \phi \), we note that the reconstruction procedure we have outlined relies only on the functional form \( P(\phi) \) without explicitly substituting its expression from Eq.~\eqref{eq:Pphi_def}. To demonstrate that this form is dynamically consistent, we evaluate the master reconstruction equation \eqref{eq:master_recon} and show in Fig.~\ref{fig:residualplot} (Appendix~A) that the residual remains small across the relevant field range. This confirms the validity of the reconstruction based on our proposed model, and we now proceed to the Einstein-frame formulation in Sec.~V.

\section{Late-Time Dynamics and Loop-Corrected Quintessence Behavior}

In this section, we examine whether the same geometric mechanism that generates early‐time inflation in our odd‐power $f(R)$ model can also drive the observed late‐time acceleration. Beginning with the Jordan–frame action, we investigate the algebraic relation between the auxiliary scalar $\varphi=f'(R)$ and the Ricci scalar to eliminate $\varphi$ and recover a pure $f(R)$ description.  We then expand the resulting field equations about the low‐curvature background $R_0$ appropriate to the present epoch.  By performing a conformal transformation to the Einstein frame, the scalar degree of freedom acquires a loop‐corrected potential which interpolates between an inflationary plateau at high curvature and a flattened, quintessence‐like form at low curvature.  Finally, we show that the late‐time dynamics of this potential are accurately captured by a thawing ansatz—i.e.\ a slowly rolling field whose equation of state departs only slightly from $w=-1$.  In what follows, we (i) carry out the low‐curvature expansion, (ii) derive the full Einstein‐frame potential including one‐loop logarithmic corrections, and (iii) compute the redshift‐dependent equation of state to assess the model’s viability as a dark‐energy candidate.

\subsection{Low-Curvature Expansion}
We begin by expanding the theory around a low, nearly constant background curvature \(R_0\), relevant to the present epoch. Denoting small fluctuations by \(R = R_0 + X\), with \(|X| \ll R_0\), we Taylor-expand the loop-corrected function \(f(R)\) as:
\begin{equation}
f(R) = f(R_0) + f'(R_0) X + \frac{1}{2} f''(R_0) X^2 + \cdots. \label{eq:fRexpand}
\end{equation}
Similarly, the first derivative becomes
\begin{equation}
f'(R) = f'(R_0) + f''(R_0) X + \cdots. \label{eq:fRprime}
\end{equation}
For notational clarity, we define background quantities as:
\begin{equation}
P_0 \equiv f'(R_0), \qquad Q_0 \equiv f(R_0) - R_0 f'(R_0). \label{eq:PQdef}
\end{equation} and denote the higher-order coefficients as
\begin{equation}
    P_1 \equiv f''(R_0), \qquad Q_1 \equiv f'(R_0) + R_0 f''(R_0).
\end{equation}

The field equations derived by varying Eq. \eqref{eq:S_Jordan} with respect to the metric \(g_{\mu \nu}\) in the Jordan frame takes the form:
\begin{equation}
f'(R) G_{\mu\nu} + \left[g_{\mu\nu} \Box - \nabla_\mu \nabla_\nu\right] f'(R) - \frac{1}{2} f(R) g_{\mu\nu} = \kappa^2 T_{\mu\nu} \label{eq:metricfR}
\end{equation}where $\Box = g^{\rho\sigma} \nabla_\rho \nabla_\sigma$ and $T_{\mu\nu}$ is the matter energy-momentum tensor. By substituting Eqs.~\eqref{eq:fRexpand}–\eqref{eq:PQdef} into Eq.~\eqref{eq:metricfR}, we obtain the following:
\begin{dmath}
    (P_0 + P_1 X) G_{\mu\nu} + \left[ g_{\mu\nu} \Box - \nabla_\mu \nabla_\nu \right](P_0 + P_1 X)
- \frac{1}{2} (2\kappa^2 Q_0 + Q_1 X)\, g_{\mu\nu} = \kappa^2 T_{\mu\nu}, \label{eq:lowcurvmetric}
\end{dmath} The constant $P_0$ can be absorbed into a redefinition of the gravitational constant \(8\pi G_{\text{eff}} \equiv \kappa^2 / P_0\) and term proportional to $Q_0$ acts as an effective cosmological constant:
\begin{equation}
\Lambda_{\text{eff}} = -\frac{Q_0}{P_0}. \label{eq:lambdaeff}
\end{equation} Meanwhile, the derivative terms acting on $X$ are suppressed by the smallness of $X$ and its slow spacetime variation. To find the leading order in this limit, one can consistently neglect higher derivatives like $\nabla_\mu \nabla_\nu X$ and $\Box X$, which are subdominant compared to the algebraic terms. Neglecting higher-order terms, Eq.~\eqref{eq:lowcurvmetric} becomes:
\begin{align}
G_{\mu\nu} + \Lambda_{\rm eff}\,g_{\mu\nu}
&= 8\pi G_{\rm eff}\,T_{\mu\nu} \nonumber \\
&\quad + \frac{f''(R_0)}{f'(R_0)}\,\bigl[g_{\mu\nu}\Box - \nabla_\mu\nabla_\nu\bigr] X
\label{eq:lowcurvEinstein}
\end{align}
where the last term in the RHS of the expression is suppressed by the small curvature deviation $X$. 

At leading order, the theory is therefore indistinguishable from GR  with a rescaled Newton’s constant and a small cosmological constant. Higher-order corrections proportional to $f''(R_0)$ only enter at order $X$ and can be neglected for sufficiently low curvature today. However, the scalar nature of \(f(R)\) gravity becomes manifest upon moving to the Einstein frame. In the coming sections, we will examine how these corrections manifest in the scalar field dynamics and analyze whether the dynamics can be consistently attributed to a slowly rolling quintessence-like behavior.
\subsection{Canonical Einstein–Hilbert form and Loop corrected Potential}
The dynamical behavior of the theory at late times, especially in relation to quintessence-like evolution, can be conveniently performed in the Einstein frame potential \cite{faraoni2004scalar, de2010f, nojiri2011unified}. This means that the gravitational sector is to be cast into canonical Einstein–Hilbert form, with all modifications encoded in a scalar field minimally coupled to gravity but non-minimally coupled to matter. For instance, if the exponential factor in $V(\phi)$ arises naturally from the $R^2$ structure in a $f(R)$ model, while the logarithmic modulation originates from one-loop corrections, one could observe that these manifestations admit a viable quintessence behavior that could match with the late-time cosmic acceleration. 

Starting from the action in Jordan frame in Eq. \eqref{eq:S_Jordan}, expressed in terms of the auxiliary field, we perform the conformal transformation to the Einstein frame:
\begin{equation}
g^E_{\mu\nu} = P(\phi) \, g_{\mu\nu}, \qquad \Omega^2(x) \equiv P(\phi), \label{eq:conformal}
\end{equation}which rescales the metric such that the Ricci scalar transforms as:
\begin{equation}
    R = \Omega^2 \left[ R_E + 3 \Box_E \ln \Omega^2 - \frac{3}{2} g^{\mu\nu}_E \partial_\mu \ln \Omega^2 \, \partial_\nu \ln \Omega^2 \right], \label{eq:56}
\end{equation}where all quantities with subscript $E$ refer to the Einstein-frame metric $g^E_{\mu\nu}$. Substituting Eq. \eqref{eq:56} into the action and integrating by parts to eliminate total derivatives yields:
\begin{align}
S_E &= \int d^4x\, \sqrt{-g_E} \bigg[
\frac{1}{2\kappa^2} R_E
- \frac{3}{4\kappa^2} \left( \frac{\partial_\mu P(\phi)}{P(\phi)} \right)^2 \nonumber \\
&\qquad\qquad
- \frac{1}{2\kappa^2} \frac{\widetilde{Q}(\phi)}{P^2(\phi)}
\bigg].
\label{eq:SEinEinstein}
\end{align}
We now (re)define a canonically normalized scalar field $\varphi$ via:
\begin{equation}
\varphi = \sqrt{\frac{3}{2}} M_{\text{Pl}} \ln P(\phi), \label{eq:phiP}
\end{equation}so that the kinetic term becomes canonical in natural units:
\begin{equation}
    \frac{3}{4\kappa^2} \left( \frac{\partial_\mu P(\varphi)}{P(\varphi)} \right)^2 = \frac{1}{2} (\partial_\mu \varphi)^2.
\end{equation} This leads to the Einstein-frame scalar field action:
\begin{equation}
S_E = \int d^4x\sqrt{-g_E}\Bigl[\tfrac{M_{\rm Pl}^2}2R_E - \tfrac12(\nabla\varphi)^2 - V(\varphi)\Bigr],
 \label{eq:SEinstein}
\end{equation}
where the scalar potential is defined by:
\begin{equation}
V(\varphi)
  = \frac{M_{\rm Pl}^2}{2\kappa^2}
    \;\frac{\widetilde Q\bigl(\phi(\varphi)\bigr)}{P^2\bigl(\phi(\varphi)\bigr)},
 \label{eq:Vphi}
\end{equation} with \(\phi(\varphi) = P^{-1} (e^{\sqrt{\frac{2}{3}}\varphi/M_{\rm pl}})\).

At this stage, the dynamics of the original $f(R)$ theory are fully recast into the Einstein frame as GR coupled to a scalar field $\varphi$ with a specific self-interaction potential $V(\varphi)$. The form of this potential is determined by the structure of $f(R)$ and through the quantum loop corrections that enter via $\widetilde{Q}(\phi)$. One could demonstrate this by taking the explicit form of $f(R)$ including the one-loop correction:
\begin{equation}
f(R) = R + \alpha R^2 + \beta R^2 \ln\left( \frac{R}{\mu^2} \right), \label{eq:fRloop}
\end{equation}
where \(\alpha, \beta\) are model-dependent constants and \(\mu\) is the renormalization scale. Differentiating Eq. \eqref{eq:fRloop} gives
\begin{equation}
    f'(R) = 1 + 2\alpha R + 2\beta R \ln\left( \frac{R}{\mu^2} \right) + \beta R. \label{eq:fderivative}
\end{equation} We note that solving for $R$ in terms of $f'(R)$, or equivalently $\phi$, is generally not possible in closed form. However, for illustrative purposes, we focus on the general behavior of the Einstein-frame potential in this limit as follows. At the classical (tree-level) level, the potential takes the exponential form \(V_{}(\varphi) \propto e^{-2\sqrt{2/3}\, \varphi/M_{\rm Pl}}.\) This arises directly from the relation $P(\phi) \sim e^{\sqrt{2/3}\, \varphi/M_{\rm Pl}}$, so that $1/P^2(\phi) \sim e^{-2\sqrt{2/3}\, \varphi/M_{\rm Pl}}$. In models such as Starobinsky inflation and no-scale supergravity \cite{linde1994hybrid}, this exponential form supports either inflation or quintessence depending on the field range and slow-roll conditions. 

In our case, a form of the potential consistent with Eq.~\eqref{eq:fderivative} is explicitly shown in Appendix C. The loop-corrected potential takes the form:
\begin{align}
   V(\varphi)
   &= V_{0}\,\exp\left[-\,\mu\,\varphi\right]\,
   \left[\,1 + \delta\,\ln\left(\tfrac{\varphi}{\varphi_{0}}\right)\right], \label{eq:92} \\
   \mu &= \frac{2\sqrt{\tfrac{2}{3}}}{M_{\rm Pl}}, \nonumber
\end{align} where $\delta \sim \beta/\alpha \ll 1$ quantifies the strength of the one-loop correction arising from vacuum polarization effects in the underlying \( f(R) \) action, and $\varphi_0$ is a renormalization scale related to the ultraviolet (UV) cutoff or the Planck scale. Physically, the tree-level exponential term governs the slow-roll evolution of the scalar field \( \varphi \) at large field values, while the logarithmic correction modifies the potential shape at subleading order. The additive loop correction in Eq.~ \eqref{eq:loopcorrpoten} can be absorbed multiplicatively into Eq.~\eqref{eq:92} under the identification \( \delta \equiv  \mathcal{Z}/U(\varphi) \ll 1 \) (as a leading-order approximation in field regions) where the loop correction is small compared to the classical potential. This leads to a relation \( \delta \sim \beta/\alpha \sim \mathcal{Z}/U(\varphi) \). Provided that \(\mathcal{Z}<0\), the sign and magnitude of \( \delta \) influence the flattening or steepening of the potential, thus affecting the scalar field’s effective equation of state. At late times, as the scalar field rolls toward larger values (and $R \to 0$), the logarithmic term becomes increasingly significant--exponential potentials become common in scalar-tensor theories, and can be modeled for thawing dark energy.

To this end, it is important to keep in mind some of the assumptions we have used in previous developments before we move to the dynamics of the model. Recall that Eq. \eqref{eq:loopcorrpoten} and Eq. \eqref{eq:92} describe the Einstein-frame potential derived from the same underlying loop-corrected \( f(R) \) gravity, but via distinct parametrizations. Equation \eqref{eq:loopcorrpoten} results from a reconstructed Jordan-frame ansatz mapped via \( \phi = e^{\sqrt{2/3}\kappa\varphi} \), yielding a leading exponential term \( U(\varphi) \sim e^{2\sqrt{2/3}\kappa\varphi} \) with an additive quantum correction \( \mathcal{Z} \ln(\varphi/\varphi_0) \). In contrast, Eq. \eqref{eq:92} follows from an effective one-loop treatment of the action, where quantum effects appear multiplicatively in \(V(\varphi)\). These forms are consistent to leading order in \( \delta \ll 1 \), since \( \delta \sim \beta/\alpha \sim \mathcal{Z}/U(\varphi) \), and differ only in their expansion scheme. Moreover, the use of distinct parametrizations is helpful to maintain analytic intractability of inverting \( f'(R) \) and constructing closed-form potentials in the Einstein frame when logarithmic corrections are present. The constant term \( (A/(2\kappa^2 \varphi_c^2)) \) in Eq. \eqref{eq:loopcorrpoten} becomes subdominant at large \( \varphi \), or can be absorbed into the normalization of \( V_0 \) without affecting the logarithmic running of the model and is therefore omitted in the effective form of \(V(\varphi)\).

\subsection{Quintessence Dynamics and Redshift-Dependent EoS}
The scalar energy density and pressure are:
\begin{equation}
\rho_\varphi = \frac{1}{2} \dot{\varphi}^2 + V(\varphi), \qquad
p_\varphi = \frac{1}{2} \dot{\varphi}^2 - V(\varphi). \label{eq:rho_p_phi}
\end{equation}
The effective equation of state is:
\begin{equation}
w_\varphi = \frac{p_\varphi}{\rho_\varphi} = \frac{\tfrac{1}{2} \dot{\varphi}^2 - V(\varphi)}{\tfrac{1}{2} \dot{\varphi}^2 + V(\varphi)}. \label{eq:wphi}
\end{equation}For late-time acceleration, the potential must dominate over the kinetic energy, i.e., $\dot{\varphi}^2 \ll V(\varphi)$. In the slow-roll regime, the EoS becomes:
\begin{equation}
    w_\varphi \approx -1 + \frac{\dot{\varphi}^2}{V(\varphi)} + \mathcal{O}\left( \frac{\dot{\varphi}^4}{V(\varphi)^2} \right). \label{eq:wphiSR}
\end{equation} Thus, $w_\varphi \to -1$ as $\dot{\varphi} \to 0$. This occurs naturally as the field rolls down an asymptotically flat or mildly decaying potential. The rate at which $w_\varphi$ approaches $-1$ depends on the flatness of the potential and the slope $V'(\varphi)$, which is determined by the exponential and logarithmic terms. Explicitly, the first derivative of the potential (Eq. \eqref{eq:92}) reads:
\begin{equation}
    V'(\varphi)
=V_{0}\,\exp[-\,\mu\,\varphi]\;
\Bigl[\,-\,\mu\,\bigl(1 + \delta\,\ln\tfrac{\varphi}{\varphi_{0}}\bigr)
\;+\;\frac{\delta}{\varphi}\Bigr].  \label{eq:Vloop}
\end{equation} The first term dominates for large $\varphi$, leading to slow-roll. The second term introduces a subleading modulation from the loop correction, slightly changing the field’s velocity and delaying the transition to full vacuum domination.

\begin{figure*}
    \centering
    \begin{subfigure}[b]{0.45\textwidth}
        \centering
        \includegraphics[width=\linewidth]{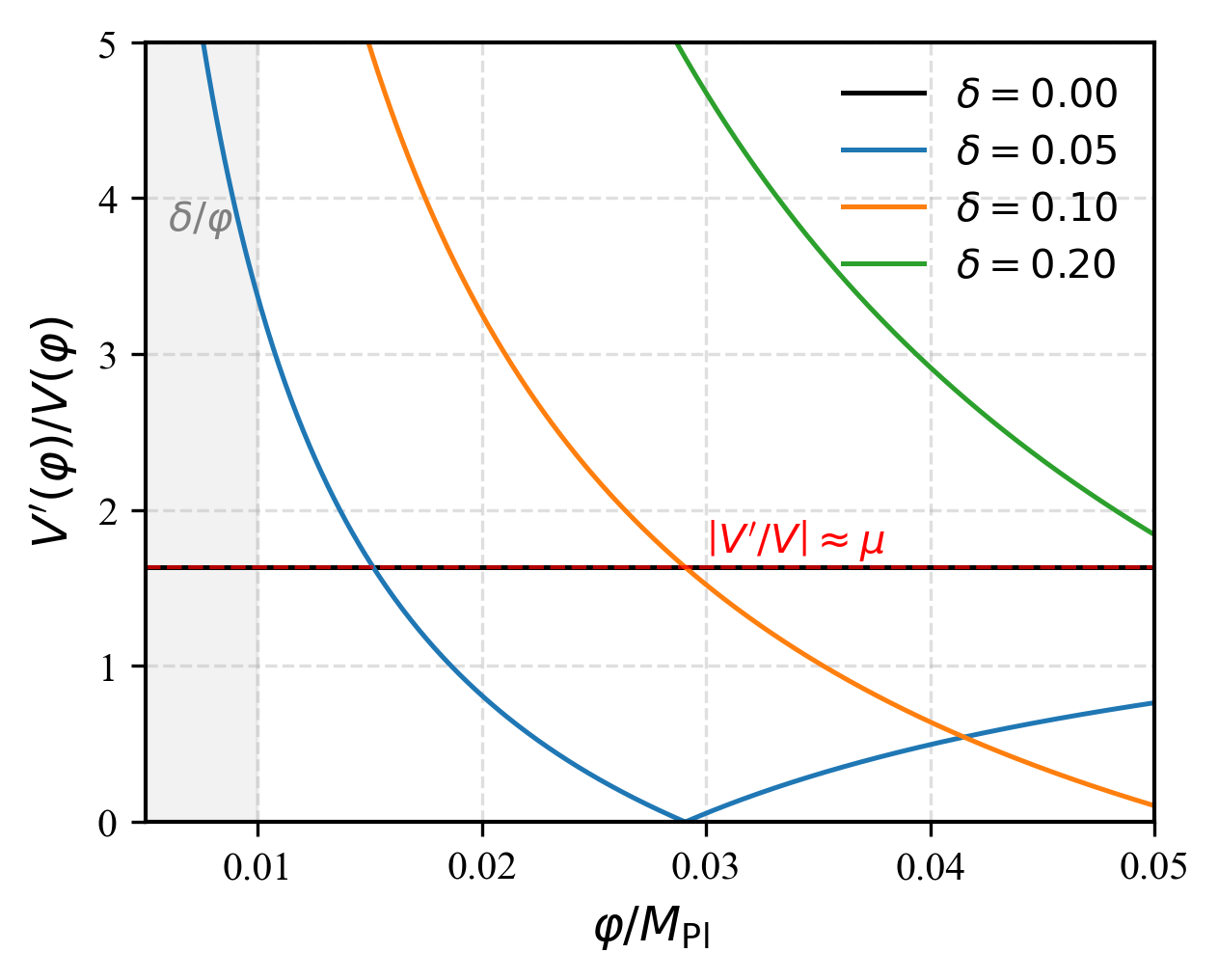}
        \caption{\small First-derivative amplitude \(\left|V'(\varphi)/V(\varphi)\right|\) vs. \(\varphi/M_{\rm Pl}\). The loop correction increases the slope at small \(\varphi\), deviating from the pure exponential.}
        \label{fig:Vprime}
    \end{subfigure}
    \hfill
    \begin{subfigure}[b]{0.45\textwidth}
        \centering
        \includegraphics[width=\linewidth]{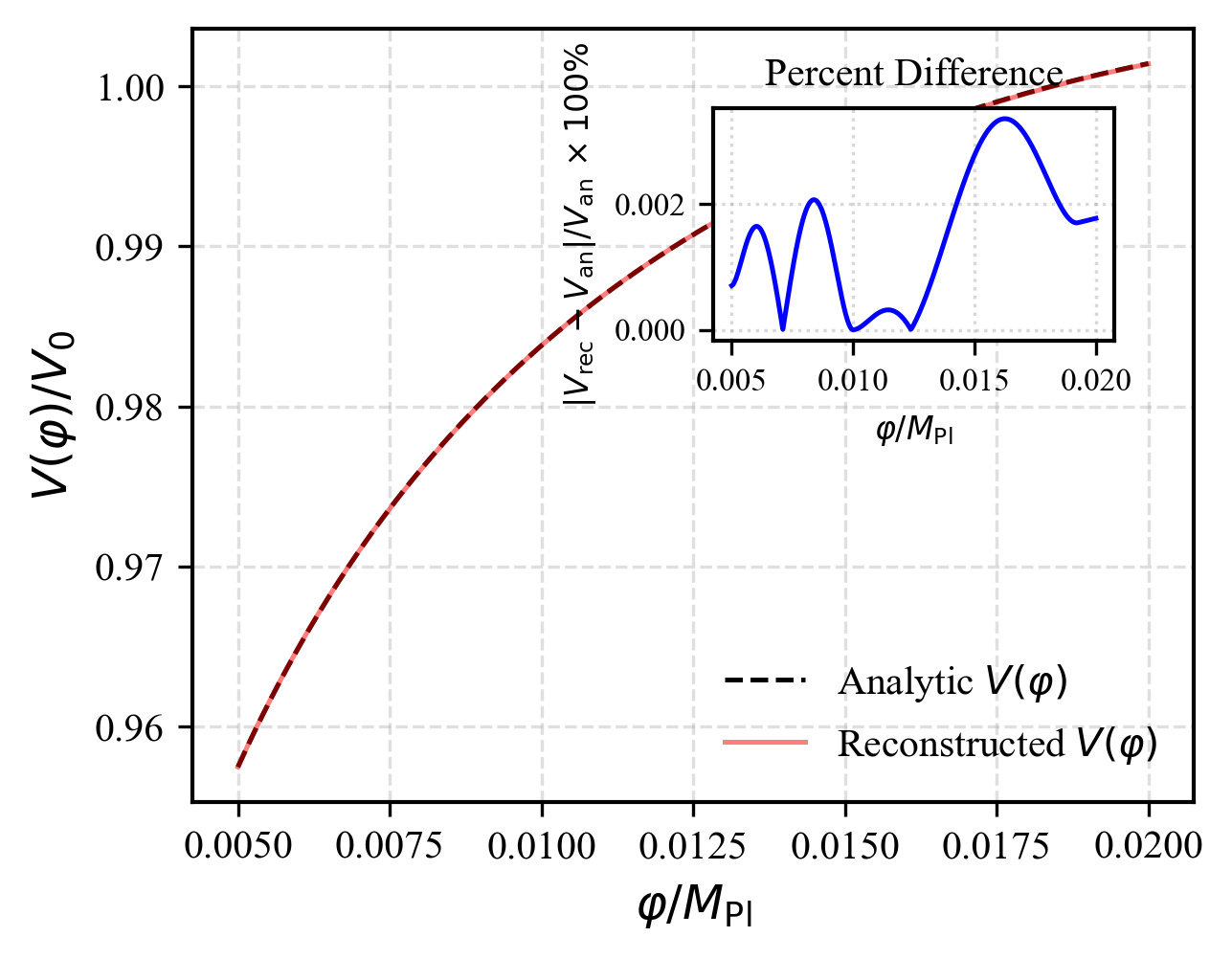}
        \caption{\small Reconstruction of the loop-corrected scalar potential $V(\varphi)$ from its curvature $\eta_{V}(\varphi)$.}
        \label{fig:Vrecon}
    \end{subfigure}
    \hfill
    \begin{subfigure}[b]{0.45\textwidth}
        \centering
        \includegraphics[width=\linewidth]{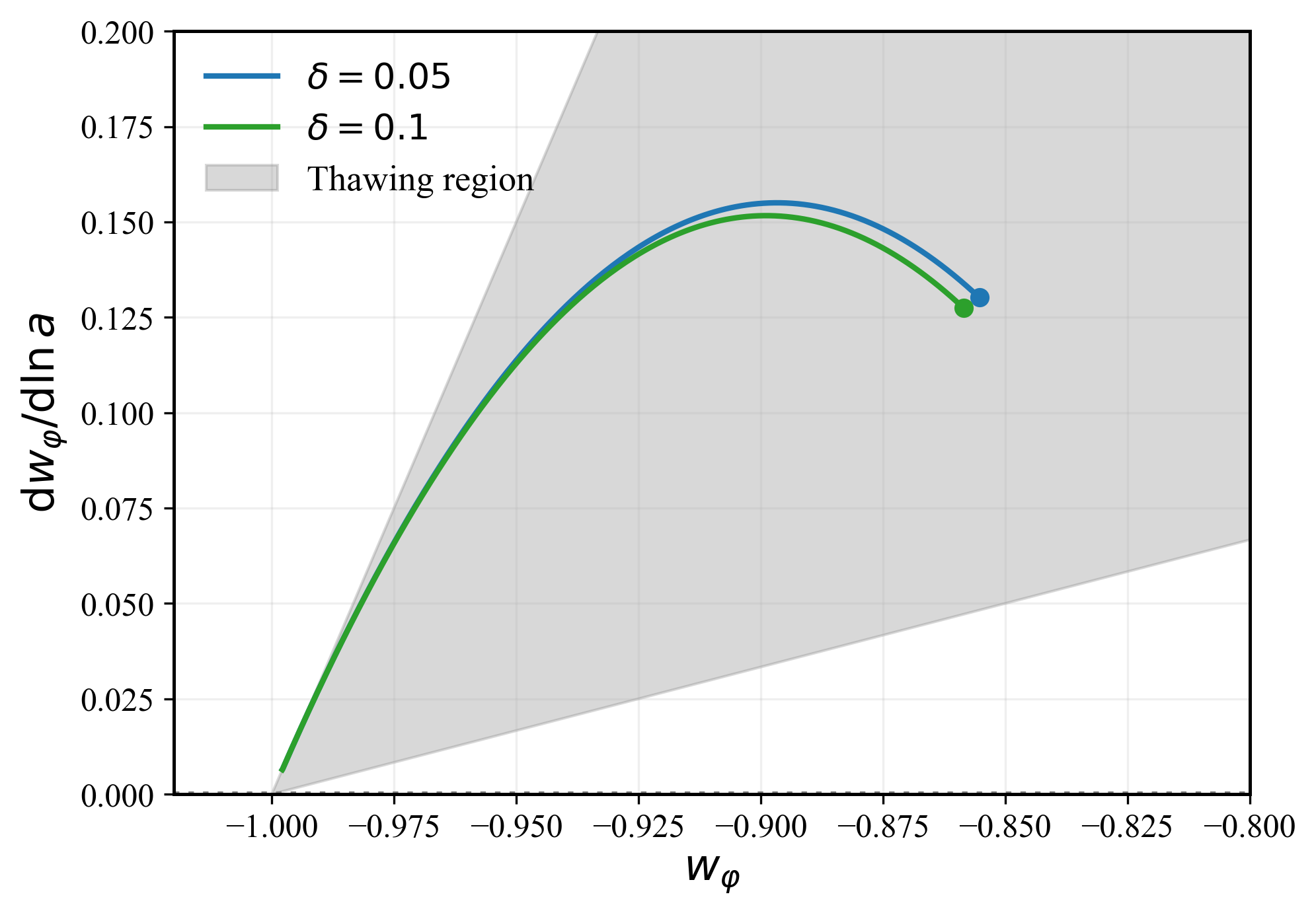}
        \caption{\small Thawing trajectory in the \((w_\varphi,\,dw_\varphi/d\ln a)\) phase space, compared to the Caldwell–Linder thawing bound \cite{caldwell2005limits}.}
        \label{fig:thawing}
    \end{subfigure}
    \caption{\small Dynamical features of the loop-corrected quintessence potential. The dashed horizontal line at $\left|V'/V\right| = \mu$ (red) marks the region where the exponential term dominates the slope. The shaded vertical band spanning $0.005 \leq \varphi/M_{\rm Pl} \leq 0.01$ (grey) highlights the domain in which the loop-correction term $\delta/\varphi$ significantly increases $\left|V'/V\right|$, deviating from $\mu$. This behavior quantifies the slow-roll parameter $\varepsilon_V$; small $\delta$ implies $\varepsilon_V \ll 1$ for $\varphi \gtrsim 0.02\,M_{\rm Pl}$, whereas larger $\delta$ introduces a steeper slope at $\varphi \lesssim 0.01\,M_{\rm Pl}$. (a) The derivative ratio \(\left|V'/V\right|\) shows enhanced slope at low \(\varphi\), critical for thawing onset. (b) The main panel shows the analytic form of $V(\varphi)/V_{0}$ (black dashed) and the numerically reconstructed $V_{\rm rec}(\varphi)/V_{0}$ (red solid) for $\varphi \in [0.005,\,0.020]\,M_{\rm Pl}$, with parameters $\varphi_{0} = 0.01\,M_{\rm Pl}$, $\lambda = 0.005\,M_{\rm Pl}$, $\delta = 0.05$, and $\mu = 2\sqrt{2/3}/M_{\rm Pl}$. The two curves coincide to within plotting precision, demonstrating that integrating $V''(\varphi)$ twice (using boundary conditions $V(\varphi_{0}) = 3\,M_{\rm Pl}^{2} H_{0}^{2} \Omega_{\Lambda,0}$ and $V'(\varphi_{0}) = -\mu\,V(\varphi_{0})$) accurately recovers the original loop-corrected potential. The inset displays the percent difference $\left|V_{\rm rec} - V_{\rm an}\right| / V_{\rm an} \times 100\%$ as a function of $\varphi / M_{\rm Pl}$, showing errors below $0.3\%$ across the entire range. (c) The thawing evolution in \((w_\varphi,\,dw_\varphi/d\ln a)\) space for \(\delta = 0.05,\,0.1\) lies entirely within the thawing wedge. The present-day point $z=0$ is marked with a circle on each curve.  Both trajectories remain confined within the thawing bound throughout the evolution, consistent with a slowly rolling field that departs from $w_\varphi = -1$ only at late times.}
    \label{fig:loopdynamics}
\end{figure*}

The scalar field obeys the Klein–Gordon equation \(\ddot{\varphi} + 3H \dot{\varphi} + V'(\varphi) = 0,\) and under slow-roll approximation ($\ddot{\varphi} \ll 3H\dot{\varphi}$, $\dot{\varphi}^2 \ll V$), this reduces to:
\begin{equation}
\dot{\varphi} \approx -\frac{V'(\varphi)}{3H}. \label{eq:phidot}
\end{equation}
Substituting Eq. \eqref{eq:phidot} into Eq.~\eqref{eq:wphiSR}, and neglecting corrections in quadratic order of \(\dot{\varphi}\), we obtain:
\begin{equation}
w_\varphi(z) \approx -1 + \frac{1}{18H^2} \left( \frac{V'(\varphi)}{V(\varphi)} \right)^2. \label{eq:wz}
\end{equation}
From the potential Eq. \eqref{eq:92} and Eq.~\eqref{eq:Vloop}, the slope can be written as:
\begin{equation}
\frac{V'(\varphi)}{V(\varphi)}
\;=\; -\,\mu\bigl(1 + \delta\,\ln\tfrac{\varphi}{\varphi_{0}}\bigr)
\;+\;\frac{\delta}{\varphi}, \label{eq:VprimeoverV}
\end{equation} to $\mathcal{O}(\delta)$ one finds \({V'(\varphi)}/{V(\varphi)} \;\approx\; -\,\mu \;+\;\frac{\delta}{\varphi},\) where we can substitute Eq. \eqref{eq:VprimeoverV} into Eq. \eqref{eq:wz} to get:
\begin{equation}
w_{\varphi}(z)
\;\approx\; -\,1 \;+\; \frac{1}{18\,H^{2}}
\Bigl[\,-\,\mu + \tfrac{\delta}{\varphi(z)}\Bigr]^{2}. \label{eq:wzfinal}
\end{equation}As \(\varphi(z)\) increases slowly, the logarithmic term decays, and \(w_\varphi(z)\) smoothly approaches \(-1\). This is shown in Fig. \ref{fig:thawing} where values $\delta\lesssim1$ yield a thawing quintessence, whereas much larger $\delta$ induces rapid field evolution. The trajectory lies in the region of the $(w_\varphi, dw_\varphi/d\ln a)$ phase space as defined by \cite{de2025thawing, dimopoulos2021quintessential} and in agreement with SNIa and BAO constraints, where the characteristic evolution of a scalar field slowly rolls from $w_\varphi \approx -1$ toward less negative values. The model thus interpolates between a thawing quintessence regime at intermediate redshift and vacuum domination at late times.

\subsection{Phenomenology of the Scalar Potential Curvature}
The curvature of the scalar potential,
\begin{align}
V''(\varphi)
&= V_0\, \exp\left[-\mu\, \varphi\right] \nonumber \\
&\quad \times \bigg[
\mu^2 \left(1 + \delta\, \ln\left(\frac{\varphi}{\varphi_0}\right)\right)
- \frac{2\mu\, \delta}{\varphi}
- \frac{\delta}{\varphi^2}
\bigg]
\label{eq:curvature}
\end{align}
governs multiple aspects of the model's cosmological viability. Most directly, the second derivative \( V''(\varphi) \) determines the effective mass of fluctuations,
\begin{equation}
m_{\rm eff}^{2}(\varphi) \equiv V''(\varphi),
\label{eq:meff}
\end{equation}
which enters both the background and perturbative dynamics.

\begin{figure*}[h]
    \centering
    \includegraphics[width=\linewidth]{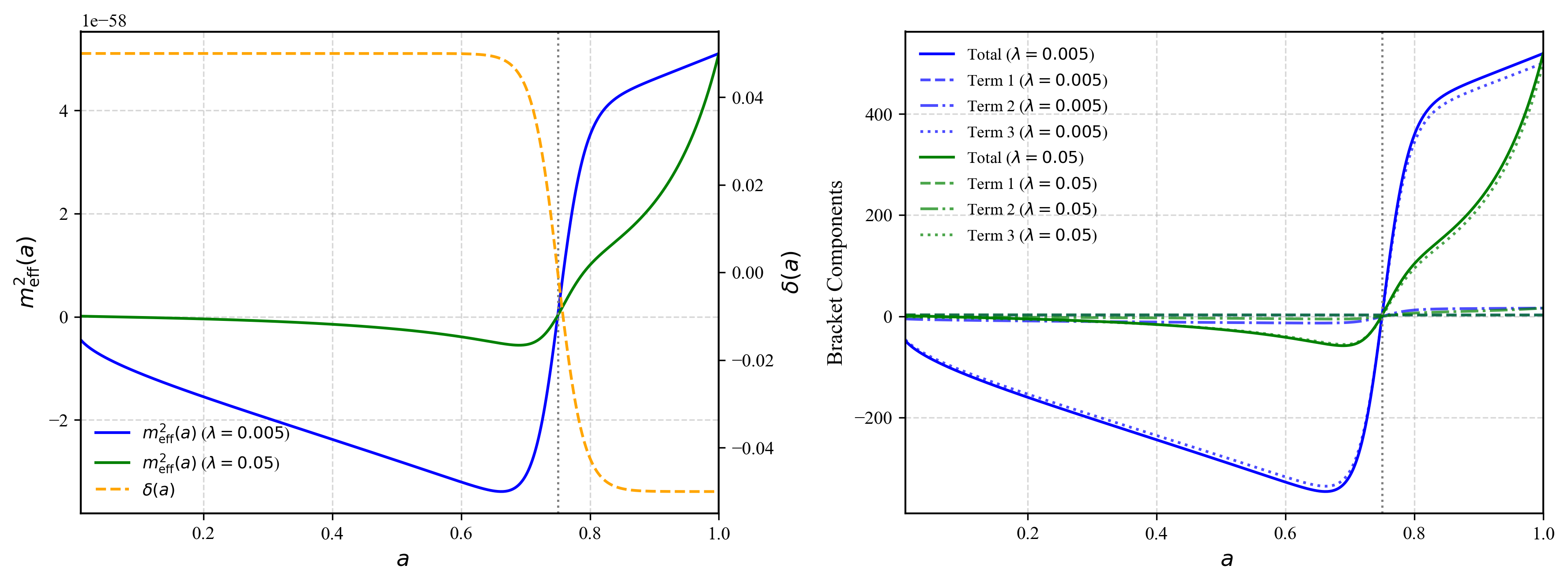}
    \caption{\small Evolution of $m_{\rm eff}^2(a)$ and the scalar potential curvature components for two thawing amplitudes $\lambda = 0.005$ and $\lambda = 0.05$, under a dynamically interpolating loop correction $\delta(a)$. The left panel shows that larger $\lambda$ accelerates the thawing onset and suppresses the late-time mass plateau. The right panel reveals how the contributions from terms inside the brackets $([\dots])$ in Eq. \eqref{eq:meffa} respond nonlinearly to $\lambda$, reshaping the effective curvature and modulating the transition (grey dotted line) from early instability to late-time stability.}
    \label{fig:thawingamplitude}
\end{figure*}

Imposing a thawing ansatz,
\begin{align}
\bar{\varphi}(z) &= \varphi_0 + \lambda \ln(1 + z) 
\quad \Longrightarrow \quad
\bar{\varphi}(a) = \varphi_0 - \lambda \ln a, \nonumber \\
\dot{\bar{\varphi}} &= -\lambda\, H(a)
\label{eq:thawing}
\end{align}
with $\lambda/M_{\rm Pl} \ll 1$, where \(\lambda\) is the field–rolling amplitude at late times. This is motivated by the fact that, in thawing quintessence models with nearly flat potentials, the Klein–Gordon equation admits quasi-analytic solutions where \(\dot{\varphi} \propto H\) (see Chiba \cite{chiba2009slow} and Dutta \cite{dutta2011slow} for a similar form). Substituting $\varphi = \bar{\varphi}(a)$ into Eq.~\eqref{eq:curvature}, one obtains:
\begin{align}
m_{\rm eff}^{2}(a)
&= V_{0} \exp\left[-\mu \left(\varphi_{0} - \lambda \ln a\right)\right] \nonumber \\
&\quad \times \bigg[
\mu^{2} \left(1 + \delta \ln\left(\frac{\varphi_{0} - \lambda \ln a}{\varphi_{0}}\right)\right) \nonumber \\
&\qquad - \frac{2\mu \delta}{\varphi_{0} - \lambda \ln a}
- \frac{\delta}{\left(\varphi_{0} - \lambda \ln a\right)^2}
\bigg]
\label{eq:meffa}
\end{align}
At early times ($a \lesssim 0.5$), the effective mass is suppressed for small $\lambda$, in order to be consistent with a frozen scalar regime. However, increasing $\lambda$ induces a faster departure from the potential plateau, thereby accelerating the onset of field rolling and raising $m_{\rm eff}^2$ earlier as shown in Figure \ref{fig:thawingamplitude} (left block). Meanwhile, decomposition of terms inside the bracket of Eq. \eqref{eq:meffa} reveals that the constituent terms are highly sensitive to $\lambda$, and regulate the mass hierarchy of the scalar sector. This is analogous to a dynamical attractor structure \cite{kazakov2023leading}, reminiscent of scale-invariant dilaton or chameleon-like models, but implemented here through purely gravitational loop corrections.

Note that one must choose $V_{0}$ such that $\mu^{2}V_{0}\sim H_{0}^{2}$ (to reproduce the observed dark-energy density today), and taking $\varphi_{0}/M_{\rm Pl} \sim \mathcal{O}(1)$, one finds numerically that $m_{\rm eff}(a)\sim H_{0}$ around $a=1$. Because $m_{\rm eff}(a)\sim H_{0}$, the Compton wavelength $\lambda_{\varphi}(a)\approx m_{\rm eff}^{-1}$ is of order the Hubble radius, and it cannot cluster on small scales. This means that the scalar field remains slowly rolling and stable under small \(\delta\) perturbations over the relevant cosmological history.  In fact, the presence of the logarithmic term leads to a dynamical adjustment of the slow-roll trajectory--the integrand in Eq. \eqref{eq:curvature} becomes more extended in field space due to a flatter effective slope, requiring smaller $\varphi_0$ values than a pure exponential potential would.  In particular, the potential slow-roll parameter:
\begin{equation}
\eta_{V}(\varphi)
= M_{\rm Pl}^{2}\,\frac{V''(\varphi)}{V(\varphi)}
\approx \frac{m_{\rm eff}^{2}}{3H^{2}} \label{eq:77}
\end{equation}
satisfies $|\eta_{V}|\ll1$ for $a\gtrsim0.5$, which is necessary to keep $w_{\varphi}$ near $-1$ until very late times.

Only once $\eta_{V}(\varphi)$ approaches unity—numerically $\eta_{V}\sim0.1$ at $z\approx0.5$ for the above parameters—does $\varphi$ roll more rapidly, producing a deviation $w_{\varphi}+1\sim\mathcal{O}(10^{-2})$. Because the kinetic term is canonical, one must have $V''(\varphi)>0$ over the field range of interest. For $\delta\in(0,\,0.1)$ and $\varphi(a)$ evolving from $\varphi_{0}$ to larger values, the bracket in Eq.~\eqref{eq:curvature} remains positive—numerically above $0.05\,M_{\rm Pl}^{-2}$ for $a\in[0.1,\,1]$. Consequently $m_{\rm eff}^{2}(a)>0$ at all times, such that the quadratic action for perturbations propagates a non-tachyonic degree of freedom. In particular, there is no ghost as long as the kinetic prefactor remains positive, which is ensured by canonical normalization in the Einstein frame. Given that the scalar Compton wavelength remains of order the Hubble radius at late times, we expect minimal impact on subhorizon clustering. Nonetheless, a full treatment involving metric perturbations and their evolution using Boltzmann codes such as CLASS or CAMB is essential to assess compatibility with large-scale structure and CMB lensing data. We leave this as a direction for future work.

The curvature $V''(\varphi)$ influences the redshift evolution of $\dot{\varphi}$ through the subleading term in the Klein–Gordon equation. Differentiating $\dot{\varphi} \approx -V'(\varphi)/(3H)$ yields
\begin{equation}
\ddot{\varphi}
\approx -\frac{V''(\varphi)\,\dot{\varphi}}{3H} + \frac{V'(\varphi)\,H'}{3H^{2}},
\label{eq:ddotphi}
\end{equation}
so that $V''(\varphi)$ directly enters second-order corrections to slow roll. Ensuring $|\ddot{\varphi}|\ll 3H\,|\dot{\varphi}|$ requires
\begin{equation}
\left|\frac{V''(\varphi)}{3H^{2}}\right|
,\quad
\left|\frac{V'(\varphi)\,H'}{3H^{3}}\right|
\ll 1. 
\end{equation}
Since $V''(\varphi)/V(\varphi)\sim \mathcal{O}(\eta_{V}/M_{\rm Pl}^{2})$ and $V'(\varphi)/V(\varphi)\sim \mathcal{O}(\mu)$, these hold for $\eta_{V}\ll1$ and $\mu\,\varphi\ll1$ at late times, given that \(\delta<0\) for \(V''(\varphi)>0\). Numerically, at $a=1$ one finds $\eta_{V}\approx0.02$ and $\mu\,\varphi\approx0.016$, so that $|\ddot{\varphi}|/(3H\,|\dot{\varphi}|)\approx0.01\ll1$.

Information regarding $V''(\varphi)$ allows partial reconstruction of the potential from observational constraints. If one obtains constraints on $\eta_{V}(z)$ or the time evolution of $w_{\varphi}(z)$, one can infer \(V''(\varphi) = m_{\rm eff}^{2}(\varphi) \equiv m_{\rm eff}^{2}(a)\) as a function of $\varphi(a)$. Integrating $V''(\varphi)$ twice with respect to $\varphi$, using boundary conditions
\begin{align}
V(\varphi_{0}) &\approx 3\,M_{\rm Pl}^{2}\,H_{0}^{2}\,\Omega_{\Lambda,0}, \label{eq:844} \\
V'(\varphi_{0}) &\approx -\mu\,V(\varphi_{0}) + \mathcal{O}(\delta).\label{eq:855}
\end{align}
yields $V(\varphi)$ up to overall constants. Eq. \eqref{eq:844} fixes the overall normalization $V_0$ once $\varphi_{0}$ and $\delta$ are known. Thus, $V_0$ is not a free parameter but is determined by matching today’s dark‑energy density. On the other hand, Eq. \eqref{eq:855} enforces the slow‑roll requirement $V'(\varphi_{0})\ll V(\varphi_{0})$, in a way that $\varphi$ today sits near the plateau and that its kinetic energy remains subdominant. In practice, one proceeds by choosing a trial $\varphi_{0}\sim\mathcal{O}(M_{\rm Pl})$ and small $\delta$, then solving for $V_0$ fixed to mimic the current value of cosmological constant \(\Lambda\).  The residual $\mathcal{O}(\delta)$ shift in $V'(\varphi_{0})$ quantifies the departure from an exact exponential plateau and determines the current value of the EoS parameter in Eq. \eqref{eq:wzfinal}.  This two‑point matching thus fully specifies the potential, removing any remaining integration freedom and tying the model directly to today’s observed cosmic acceleration. 

Although the present analysis is confined to the late-time thawing regime, a complete reconstruction of the model across all cosmological epochs would require computing the number of inflationary e-foldings to verify that the same loop-corrected potential can consistently sustain both early- and late-time acceleration, as discussed in Section II. This reconstruction is also essential for a consistent treatment of scalar perturbations in the loop-corrected $f(R)$ framework. The loop correction softens \(m_\phi\) at \(R \approx R_0\), allowing the scalaron to remain light and source dark energy. Without this, the odd-power ansatz alone would yield \(m_\phi \to \infty\) at \(R = R_0\), preventing late-time dynamics. As the scalaron acquires a canonical mass $m_\phi^2 \equiv {d^2 U(\phi)}/{d\phi^2}$ of order $\mathcal{O}(H_0)$ at late times, it mediates a scale-dependent modification to the gravitational coupling. In particular, the growth of matter overdensities is governed by an effective Newton constant $G_{\rm eff}(k,a)$ that departs from GR on scales where $k \gtrsim a m_\phi$, leading to enhanced structure formation. A full numerical analysis of this effect, incorporating the scalaron's evolution into Einstein–Boltzmann solvers such as \texttt{CLASS} or \texttt{CAMB}, would enable a joint confrontation of the model with both background and perturbation observables, placing tighter constraints on the potential parameters $(\phi_0, \lambda, Z)$. We leave such an analysis to future work.

\section{Analysis}
\begin{figure}[h]
  \centering

  \begin{subfigure}[b]{0.45\textwidth}
    \centering
    \includegraphics[width=\textwidth]{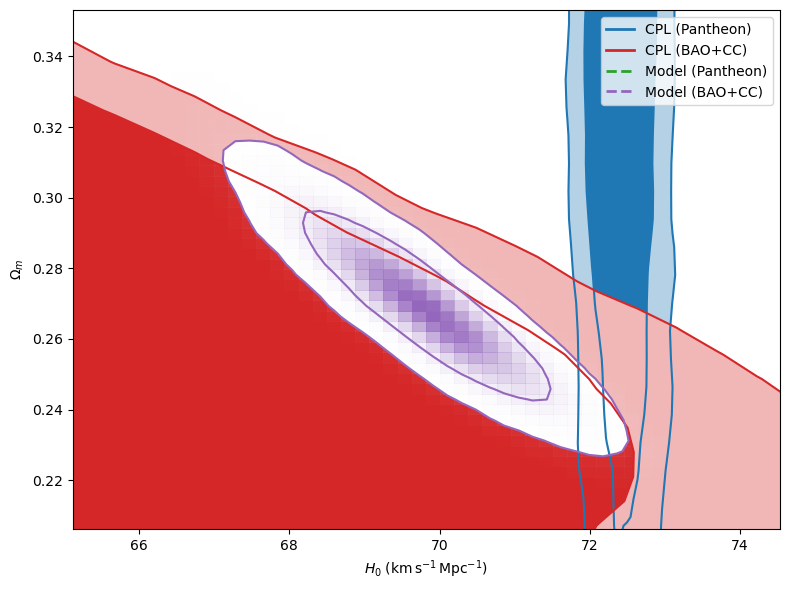} 
    \caption{$H_0$ vs. $\Omega_m$.}
    \label{fig:BAO_CC}
  \end{subfigure}
  \hfill
  \begin{subfigure}[b]{0.45\textwidth}
    \centering
    \includegraphics[width=\textwidth]{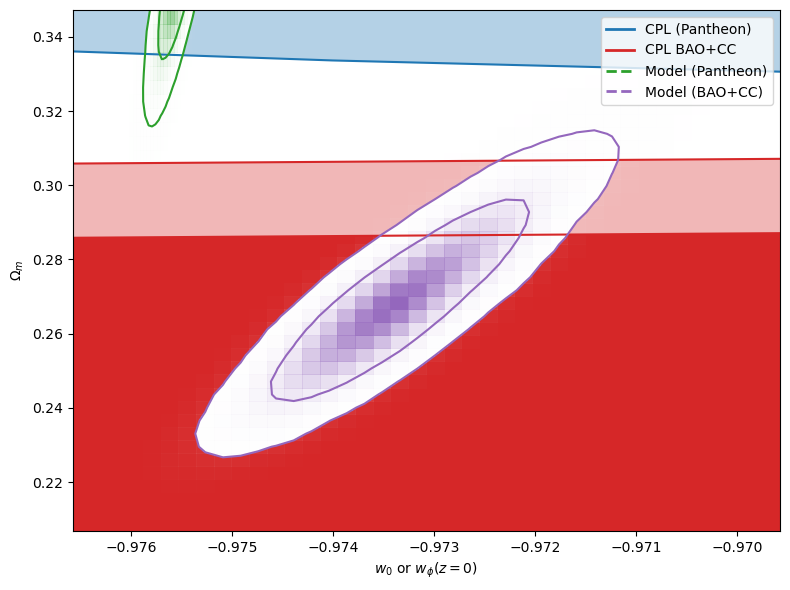} 
    \caption{$w_0 \equiv w_\varphi(z=0)$ vs. $\Omega_m$.}
    \label{fig:Pantheon_SH0ES}
  \end{subfigure}

  \caption{\small 
    Posterior distributions of cosmological parameters obtained by comparing a scalar dark energy model with the phenomenological CPL parameterization, using joint constraints from BAO + cosmic chronometer (CC) and Pantheon + SH0ES data. We compare $H_0$ and $w_0 \equiv w_\varphi(z=0)$ as functions of $\Omega_m$ to examine degeneracies and physical implications in both models. Each panel shows 68\%  and 95\%  confidence contours for the relevant parameter combinations. The scalar field EOS is derived from a slow-varying logarithmic field in Eq. \eqref{eq:thawing},  with $\varphi_0 = 0.01$, $\lambda = 1$ (set to maximum), and $\delta = 0.5$. MCMC analysis was performed using flat priors: $50 < H_0 < 90$, $0.1 < \Omega_m < 0.5$ for the scalar model, and $50 < H_0 < 73$, $0.1 < \Omega_m < 0.5$, $-2 < w_0 < -0.3$, $-2 < w_a < 2$ for the CPL model. The custom scalar field dynamics yield a self-consistent Hubble parameter evolution through iterative evaluation of the Friedmann equation, contrasting with the parametric form of CPL.
  }
  \label{fig:H0_vs_Omegam}
\end{figure}

We begin by analyzing the posterior distributions of the key model parameters $\{\lambda, V_0, \varphi_0\}$, derived from the reconstruction based on Eq. \eqref{eq:92}, and are summarized in Table \ref{tab:scalar_combined}. These posteriors constrain the thawing slope $\lambda$, the scalar field normalization $\varphi_0$, and the potential amplitude $V_0$, which together determine the redshift evolution of the effective dark energy EoS $w_\varphi(z)$. In this analysis, we assume that the scalar field evolves self-consistently on the loop-corrected potential defined in Eq.\eqref{eq:92}, and the corresponding EoS is computed using Eq. \eqref{eq:rho_p_phi} and Eq. \eqref{eq:wzfinal}. The term $(R - R_0)^{2n+1}$ in the original odd-power $f(R)$ ansatz of Eq.~\eqref{eq:fR_ansatz} induces an asymptotically exponential potential in the Einstein frame, with logarithmic loop corrections derived from the non-linear structure of $f'(R)$. The potential governs the evolution of $\varphi(z)$, which in turn determines the dynamics of $w_\varphi(z)$ and the Hubble parameter $H(z)$ via the Friedmann equations. The model exhibits thawing quintessence behavior, with $w_\varphi(z)$ evolving slowly from a nearly frozen state at high redshift to mild departures from $-1$ at late times. The allowed parameter space, particularly for $\delta \in [-1,1]$ and $\lambda \sim 0.005\,M_{\rm Pl}$, leads to trajectories that are well approximated by Chevallier–Polarski–Linder (CPL)-like behavior in the redshift range $0 \lesssim z \lesssim 2$, with $|1 + w_0| \lesssim 0.05$ and small $w_a$. The corresponding MCMC constraints are visualized in Figure \ref{fig:H0_vs_Omegam}, which shows the degeneracies between the Hubble constant, matter density, and EoS parameters under this thawing reconstruction.

\begin{table*}[h]
  \centering
  \begin{minipage}{\linewidth}
    \centering
    \begin{tabular}{lcc}
      \toprule
      \textbf{Parameter} & \textbf{Pantheon‐only} & \textbf{Pantheon+BAO+CC} \\
      \midrule
      $H_{0}$ [km\,s$^{-1}$\,Mpc$^{-1}$] & $72.96 \pm 0.22$ & $73.88 \pm 0.12$ \\
      $\Omega_{m}$                      & $0.361 \pm 0.017$ & $0.2525 \pm 0.0023$ \\
      $V_{0}\,[M_{\rm Pl}^{4}]$         & $(5.47 \pm 2.60)\times 10^{4}$ & $(5.47 \pm 2.59)\times 10^{4}$ \\
      $\varphi_{0}\,[M_{\rm Pl}]$       & $0.0274 \pm 0.0130$ & $0.0273 \pm 0.0130$ \\
      $\lambda\,[M_{\rm Pl}]$           & $0.0105 \pm 0.0055$ & $0.0103 \pm 0.0055$ \\
      \bottomrule
    \end{tabular}
  \end{minipage}%
  \hfill
  \begin{minipage}{\linewidth}
    \centering
    \begin{tabular}{c ccc ccc}
      \toprule
      $z$ 
        & \multicolumn{3}{c}\textbf{Pantheon‐only} 
        & \multicolumn{3}{c}\textbf{Pantheon+BAO+CC} \\
      \cmidrule(lr){2-4} \cmidrule(lr){5-7}
        & $w_{\varphi,\mathrm{med}}$ & $w_{\varphi,\mathrm{lo68}}$ & $w_{\varphi,\mathrm{hi68}}$
        & $w_{\varphi,\mathrm{med}}$ & $w_{\varphi,\mathrm{lo68}}$ & $w_{\varphi,\mathrm{hi68}}$ \\
      \midrule
      0.0 & $-0.997$ & $-1.000$ & $-0.982$ & $-0.997$ & $-1.000$ & $-0.983$ \\
      0.5 & $-0.999$ & $-1.000$ & $-0.994$ & $-0.999$ & $-1.000$ & $-0.993$ \\
      1.0 & $-1.000$ & $-1.000$ & $-0.998$ & $-0.999$ & $-1.000$ & $-0.997$ \\
      \bottomrule
    \end{tabular}
  \end{minipage}
  \caption{\small {Top:} Posterior constraints on the scalar‐field model parameters from Pantheon-only and Pantheon+BAO+CC datasets. While BAO+CC tighten $H_0$ and $\Omega_m$, other parameters remain largely unchanged. {Bottom:} Reconstructed $w_{\varphi}(z)$ at three redshifts for Pantheon‐only versus Pantheon+BAO+CC fits for thawing quintessence model. Each block lists the median and $68\%$ credible bounds $[w_{\varphi,\mathrm{lo68}},\,w_{\varphi,\mathrm{hi68}}]$. At $z=0$, both datasets yield nearly identical bounds.  At $z=0.5$ and $z=1.0$, the medians remain only a slight tightening of the upper edge when BAO+CC are added.  This confirms that even after including BAO+CC, the scalar‐field EoS remains effectively indistinguishable from $w_{\varphi}=-1$ (to $\mathcal{O}(10^{-3})$) over $0\le z\le1$.
  }
  \label{tab:scalar_combined}
\end{table*}

The loop-corrected potential of Eq.~\eqref{eq:92} thus features an asymptotically flat plateau for large $\varphi$, such that $\dot{\varphi}\approx0$ over a broad redshift range ($z\gtrsim1$). As $H(z)$ decreases, $\varphi$ thaws from this frozen state and slowly descends the logarithmically corrected tail of $V(\varphi)$. Consequently, $w_{\varphi}$ departs gently from $-1$ at $z\lesssim1$, producing the thawing quintessence behavior with $w_{\varphi}+1\sim\mathcal{O}(10^{-2})$. The flattening of the potential indicates attractor-like behavior, where a wide range of initial conditions converge to the slow-roll trajectory. Comparison with the CPL parametrization in Fig.~\ref{fig:H0_vs_Omegam} shows that for $\delta\lesssim0.1$ and $\lambda/M_{\rm Pl}\approx0.005$, the model satisfies observational bounds $|1+w_{\varphi}(0)|\lesssim0.05$ while maintaining a fully dynamical dark-energy sector. This represents a well-motivated realization of loop-corrected scalar-field dark energy consistent with observational constraints; however, the corresponding analysis does not favor an evolving dark energy sector over a cosmological constant.

We also analyze a simplified scenario in which the scalar field evolves according to the phenomenological thawing ansatz given in Eq. \eqref{eq:thawing}, while the background expansion history $H(z)$ is fixed to the standard $\Lambda$CDM form specified by $\{H_0, \Omega_m\}$. We decouple the scalar dynamics from backreaction effects, thereby isolating the influence of the field evolution on the observed distance–redshift relation. Two independent log-likelihoods are constructed for this analysis: one based on Pantheon+SH0ES Type Ia supernova data, and another based on a combined dataset of BAO and Cosmic Chronometer (CC) $H(z)$ measurements. In both cases, the theoretical distance modulus $\mu(z)$ is computed directly from the scalar-field evolution, allowing a direct comparison with observational constraints. Additionally, we extract the effective scalar mass as defined in Eq.\eqref{eq:meffa}, which quantifies the redshift-dependent curvature of the potential. This decomposition allows the evolution of field stability and slow-roll behavior in the thawing regime, even within a fixed background cosmology.

\begin{table*}[h]
\centering

\begin{tabular}{lcc}
\toprule
\textbf{Parameter} & \textbf{Pantheon+SH0ES} & \textbf{Pantheon+BAO+CC} \\
\midrule
$H_0$ [km\,s$^{-1}$\,Mpc$^{-1}$]     
  & $72.95 \pm 0.22$ 
  & $74.14 \pm 0.16$ \\
$\Omega_m$ 
  & $0.3615 \pm 0.018$ 
  & $0.235 \pm 0.0069$ \\
$V_0$ [$M_{\rm Pl}^{4}$] 
  & $(5.038 \pm 3.4)\times 10^{-51}$ 
  & $(4.967 \pm 3.4)\times 10^{-51}$ \\
$\varphi_0$ [$M_{\rm Pl}$] 
  & $0.0502 \pm 0.034$ 
  & $0.0497 \pm 0.034$ \\
$\lambda$ [$M_{\rm Pl}$] 
  & $0.0487^{+0.035}_{-0.033}$ 
  & $0.0506^{+0.033}_{-0.035}$ \\
\bottomrule
\end{tabular}
\caption{\small Posterior constraints on thawing scalar-field model parameters from Pantheon+SH0ES and Pantheon+BAO+CC datasets. Median values and 68\% credible intervals for each parameter are printed alongside.}
\label{tab:scalar_posterior}
\end{table*}

The posterior distributions for the free parameters $(H_0, \Omega_m, V_0, \varphi_0, \lambda)$ reveal significant differences between the two data combinations with flat priors and delta fixed at a small fiducial value ($\delta = 0.05$) to mimic mild loop-level corrections. Best fit parameters obtained from the analysis are shown in Table \ref{tab:scalar_posterior}. The larger Hubble constant $H_0$ and smaller matter density $\Omega_m$ preferred by BAO+CC suggest a preference for a slightly more dynamical scalar sector relative to a pure $\Lambda$CDM baseline. The inferred $V_0 \sim \mathcal{O}(10^{-51})\ M_{\rm Pl}^4$ is consistent with the observed late-time dark energy scale, while the small values of $\lambda$ and the logarithmic field amplitude $\varphi_0 \ll M_{\rm Pl}$ ensure the field evolves slowly, maintaining $w_\varphi(z) \approx -1$ with mild thawing. This means loop-corrected thawing scalar models can fit current data while alleviating fine-tuning in the initial conditions, as the logarithmic correction softens the potential curvature and delays field evolution. Note that the parameter $V_0$ in Table \ref{tab:scalar_combined} represents a self-consistent reconstruction of the scalar potential in the Einstein frame, where quantities are rescaled in reduced Planck units and the potential is expressed in a dimensionless form to match the dynamics of the thawing scalar field under slow-roll conditions. In contrast, Table \ref{tab:scalar_posterior} has results from a phenomenological thawing ansatz imposed on a fixed background.
\begin{figure*}
  \centering
  
  \begin{subfigure}[b]{0.45\linewidth}
    \includegraphics[width=\linewidth]{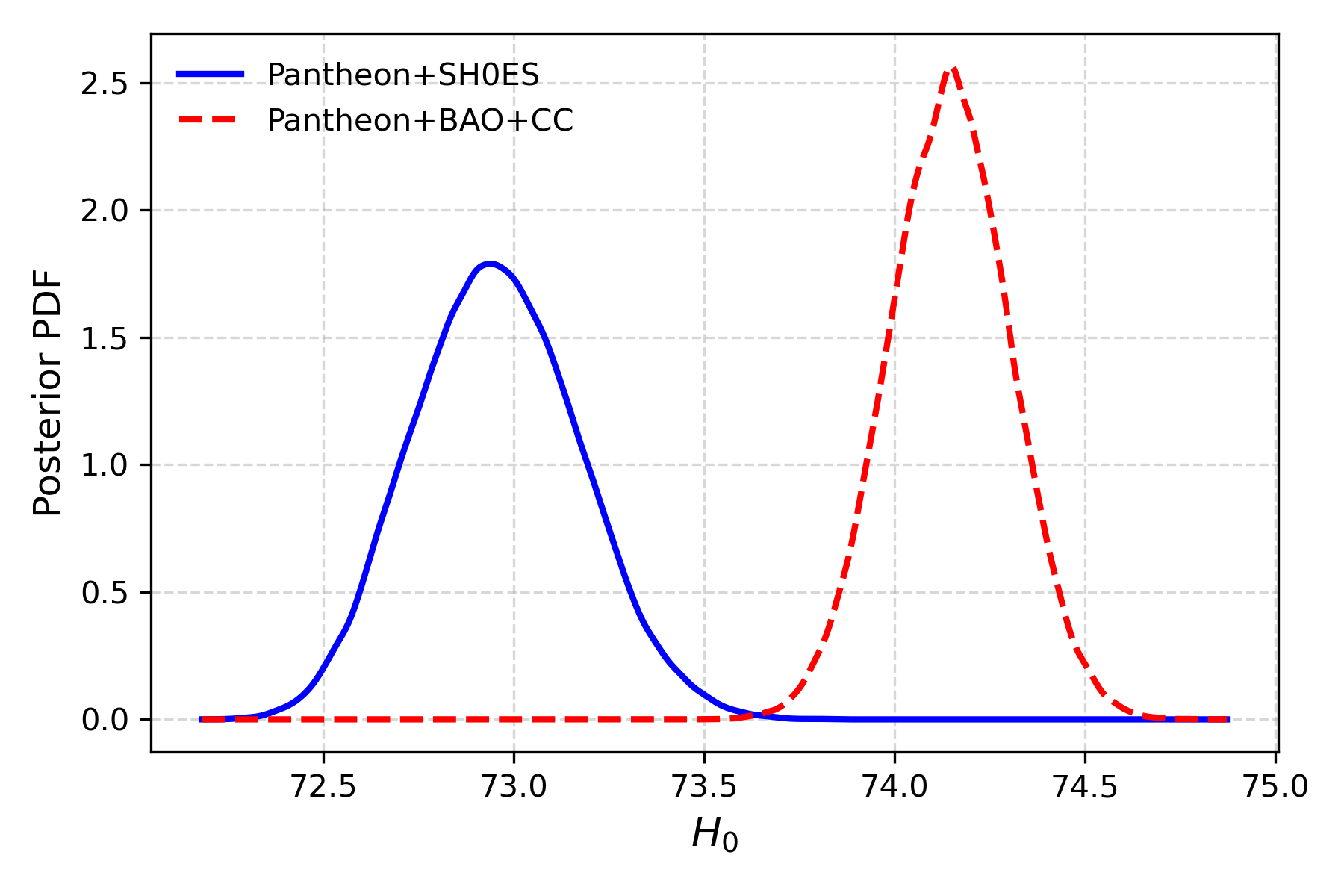}
    \caption{\small Posterior of \(H_0\). }
    \label{fig:sub1}
  \end{subfigure}
\hfill
  \begin{subfigure}[b]{0.45\linewidth}
    \includegraphics[width=\linewidth]{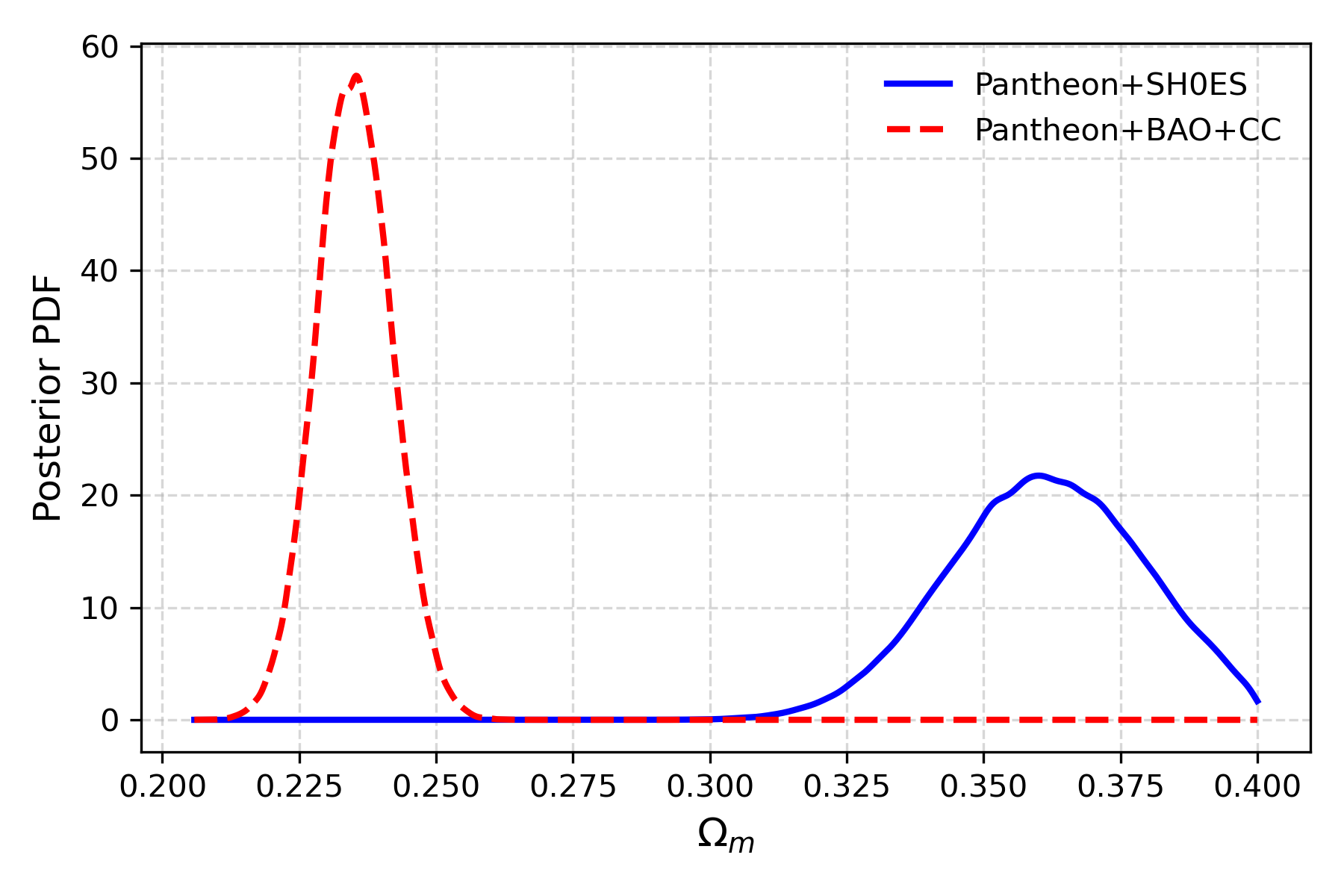}
    \caption{\small Posterior of \(\Omega_m\).}
    \label{fig:sub2}
  \end{subfigure}
\hfill
  \begin{subfigure}[b]{0.45\linewidth}
    \includegraphics[width=\linewidth]{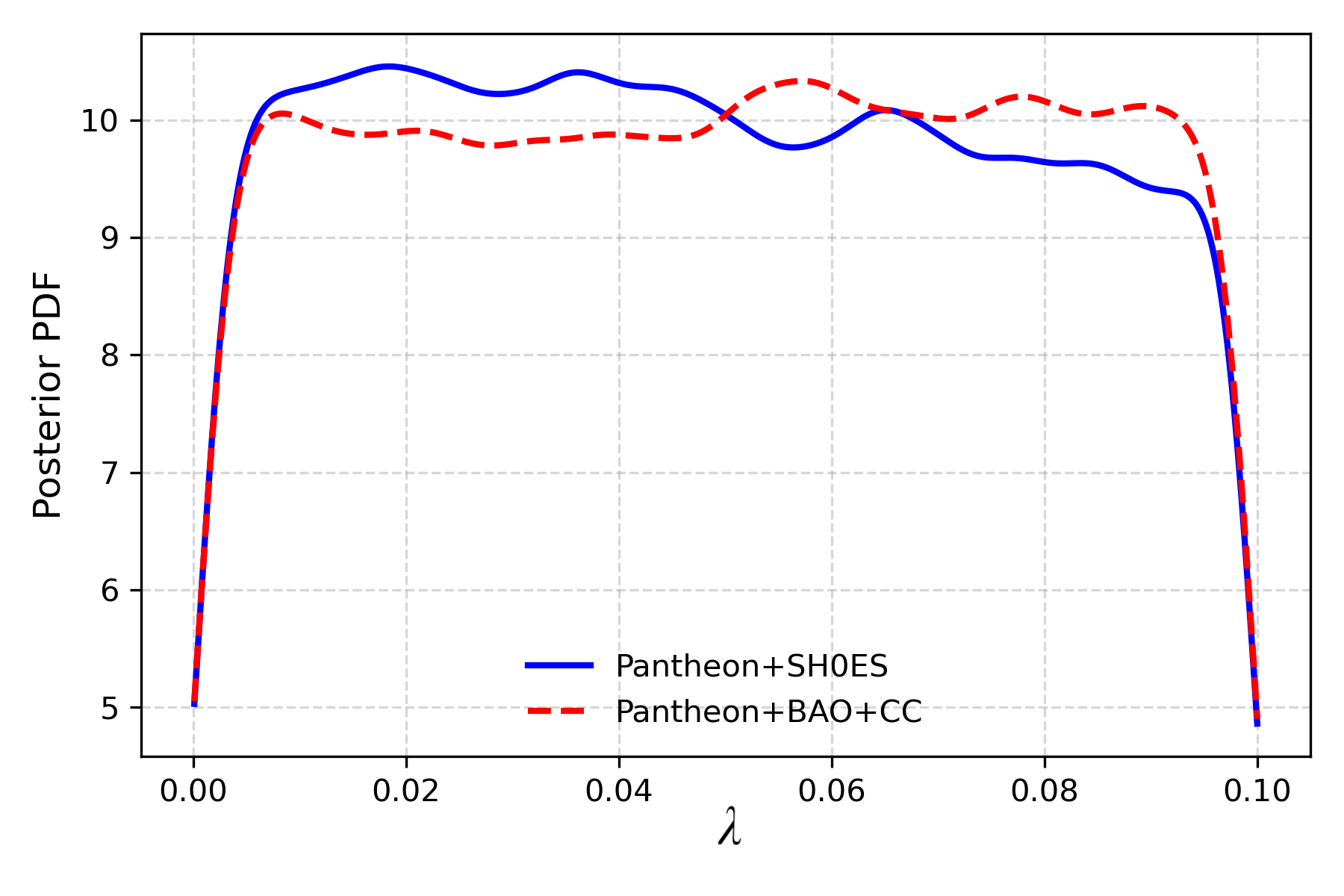}
    \caption{\small Posterior of \(\lambda\).}
    \label{fig:sub3}
  \end{subfigure}
\hfill
  \begin{subfigure}[b]{0.45\linewidth}
    \includegraphics[width=\linewidth]{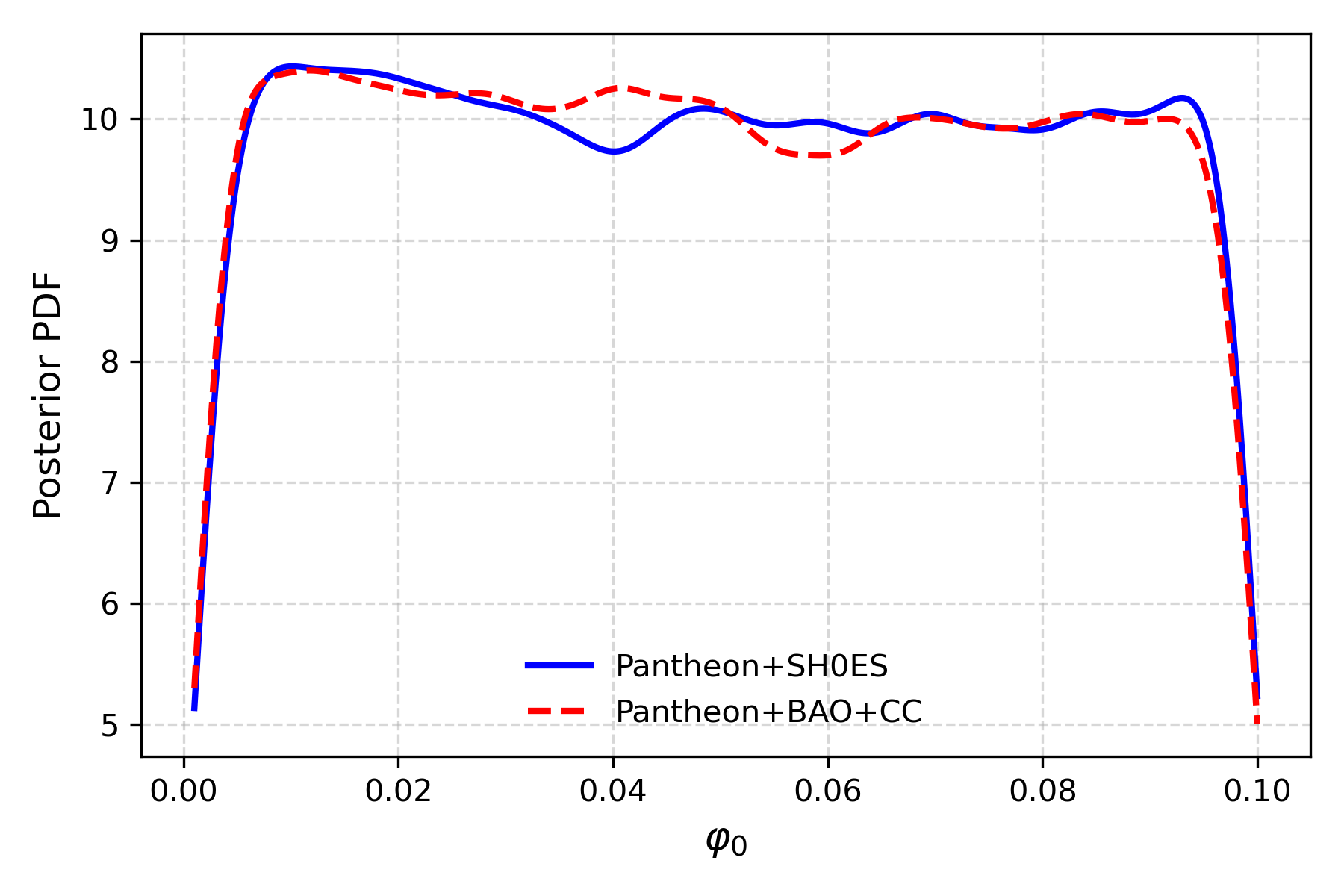}
    \caption{\small Posterior of \(\varphi_0\).}
    \label{fig:sub4}
  \end{subfigure}
  \hfill
  \begin{subfigure}[b]{0.45\linewidth}
    \includegraphics[width=\linewidth]{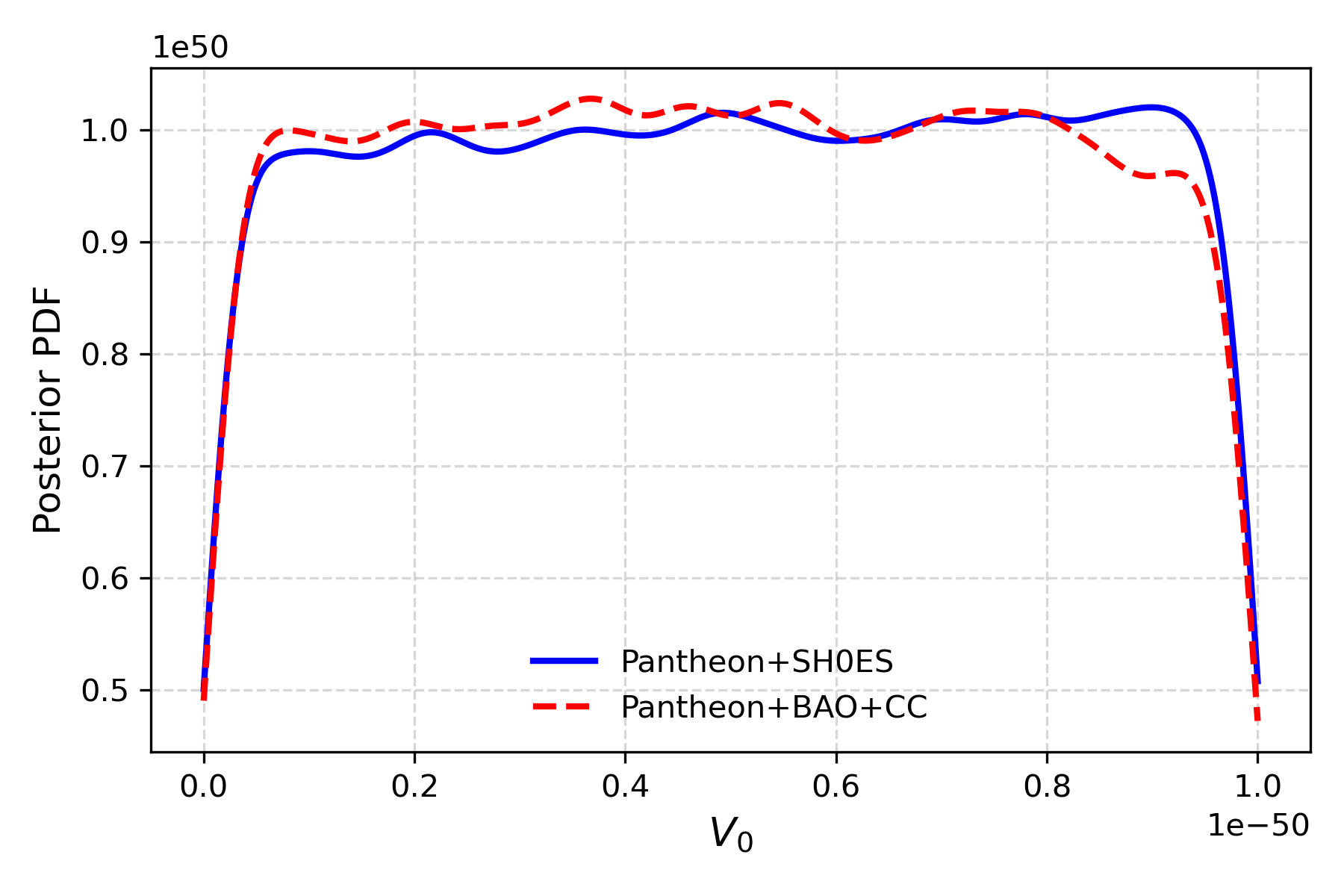}
    \caption{\small Posterior of \(V_0\).}
    \label{fig:sub5}
  \end{subfigure}
  
  \caption{\small Posterior probability distributions for each model parameter inferred from two different data combinations: The KDE reveals that the inclusion of BAO and cosmic chronometer (CC) data shifts and tightens constraints relative to supernova-only data. In particular, BAO+CC data systematically prefer a higher $H_0$, lower $\Omega_m$, and sharper localization in $V_0$ (when treated as a free parameter), implying stronger constraints on the scalar field dynamics.}
  \label{fig:five_subfigures}
\end{figure*}

While the reconstructed scalar dynamics exhibit behavior consistent with a slowly thawing field, the lack of statistically significant deviation from $w=-1$ means observational preference over $\Lambda$CDM remains marginal. Nevertheless, the upward shift in $H_0$ in the BAO+CC dataset may suggest a partial alleviation of the Hubble tension, though a dedicated analysis is needed. When the loop correction is included, the potential is flattened, which drives $w_0$ closer to $-1$; this effect is compensated in the fits by a modest increase in $\Omega_m$ and a slight downward adjustment in $H_0$, both well within current $1\sigma$ uncertainties. Thus, the loop term primarily reshuffles the late-time parameters $w_0$, $H_0$, and $\Omega_m$, while its impact on other cosmological quantities is minimal given present data, as the scalar degree of freedom does not cluster on sub-horizon scales. Future datasets from LSST and Euclid could enhance the sensitivity of dynamical dark energy models by probing the growth rate and lensing signal. Importantly, quantifying the impact of scalar perturbations on the matter power spectrum and CMB lensing is needed to distinguish this scenario from $\Lambda$CDM, although a full treatment of non-linear corrections remains to be developed.

\section{Discussion}
In this paper, we investigate a framework for unifying cosmic inflation and dark energy within metric $f(R)$ gravity, motivated by key limitations of Palatini-inspired approaches. In particular, Palatini models often suffer from decoupling of the Ricci scalar from the connection in the radiation-dominated era ($R \to 0$), rendering the theory dynamically inactive at late times. To address this, we introduce two structural assumptions within the metric formulation. First, we adopt an odd-power curvature ansatz of the form $f(R) \propto (R - R_0)^{2n+1}$, which generates a Starobinsky-like inflationary plateau at high curvature and a stable de Sitter minimum at $R = R_0$. Second, we introduce loop corrections that dynamically flatten the scalar potential $V(\varphi)$ in the infrared regime, where the scalar degree of freedom remains significant throughout cosmic evolution. The resulting Jordan-frame hilltop potential $V(\phi)$, upon conformal transformation, leads to an Einstein-frame potential corrected by a logarithmic term of the form $\delta \ln(\varphi / \varphi_0)$, which preserves inflationary behavior while enabling late-time thawing dynamics. While this additive form captures the essential flattening near the present epoch, it is analytically advantageous to recast it as a multiplicative correction, as in Eq. \eqref{eq:92}, which accurately approximates the slow-roll regime and allows closed-form solutions for scalar field evolution.

The loop correction dynamically regulates the curvature of the scalar potential $U''(\varphi)$, thereby controlling both the mass scale and evolution of the scalaron field in the dark energy era. In this framework, the effective dark energy EoS parameter $w_\varphi(z)$ becomes a sensitive probe of quantum flattening effects, as the logarithmic deformation introduced by the loop correction suppresses the steepness of the potential at low curvature. This suppression is closely tied to the odd-power index $n$ in Eq.~\eqref{eq:fR_ansatz}, which governs the smoothness of the transition between high- and low-curvature regimes. Larger values of $n$ yield a sharper transition, resulting in a steeper Einstein-frame potential upon conformal mapping. This in turn enhances both the effective slope $\lambda \sim V'/V$ and the loop amplitude $\delta \sim \beta/\alpha$, accelerating the thawing behavior of the scalar field. As a result, the present-day EoS $w_0$ is expected to deviate from the cosmological constant limit $w = -1$, and the late-time expansion rate governed by $H_0$ can shift toward higher values. These deviations are broadly consistent with low-redshift observations hinting at mild departures from $\Lambda$CDM \cite{giare2024robust}, and they offer a compelling realization of dynamical dark energy within loop-corrected modified gravity. In particular, the scalaron’s thawing rate, controlled by the loop-corrected potential, leaves imprints in both the EoS trajectory and the scalaron mass scale—features that may be tested with upcoming cosmological surveys \cite{tada2024quintessential}.

A subtle but important aspect of our framework concerns the question of frame equivalence. The hilltop potential arises in the Jordan frame as a classical geometric feature of the underlying $f(R)$ structure, while the loop correction is introduced in the Einstein frame as a low-energy quantum modification. Although these two formulations are related via a conformal transformation, they are not dynamically equivalent once quantum effects are included. In particular, the conformal mapping does not preserve the full action-level structure nor ensure the invariance of physical observables such as particle masses and scalar perturbations. Loop-induced terms typically break conformal symmetry, rendering the Einstein frame more appropriate for phenomenological analysis. Nevertheless, the essential dynamical ingredient—the scalaron—remains the same, arising purely from the curvature structure of the theory without the need for additional fields or nonminimal couplings. For moderate loop amplitudes $|\delta| \ll 1$, the scalar field remains stabilized across radiation and matter domination, preserving consistency with nucleosynthesis and inflationary constraints. Importantly, this setup allows for thawing behavior at late times without invoking new phase transitions or finely tuned potentials. As previously emphasized in the context of hilltop quintessence models \cite{scherrer2008thawing, dutta2008hilltop}, the scalaron naturally satisfies slow-roll conditions like Eq. \eqref{eq:77}, yielding viable present-day values for $w_\phi$ and the dark energy density $\Omega_\phi$ within observational bounds.

Consistent with the theoretical expectations discussed above, constraints from Pantheon+SH0ES and BAO+CC datasets reveal no statistically significant evidence for evolving dark energy. The best-fit scalar field amplitude $\varphi_0 = 0.0274 \pm 0.0130\,M_{\rm Pl}$ and thawing slope $\lambda = 0.0105 \pm 0.0055\,M_{\rm Pl}$ yield an equation-of-state trajectory characterized by $w_\varphi(0) = -0.997$ and $w_\varphi(1) = -1.000 \pm 0.002$, effectively indistinguishable from $\Lambda$CDM at the $10^{-3}$ level (see Table \ref{tab:scalar_combined}). We reconstruct this trajectory using the CPL parametrization, finding that the model occupies the same narrow region in $w_0$–$w_a$ space as other thawing scenarios. In both Pantheon-only and combined fits, the loop correction amplitude $\delta$ remains prior-driven, exhibiting no statistically favored deviation from zero. By constraining the scalar potential parameters $(V_0, \delta)$ and field amplitude parameters $(\varphi_0, \lambda)$, we find that the field remains nearly frozen across the redshift range $z \in [0, 1]$, with minimal deviation from $\Lambda$CDM. These results suggest that while the thawing potential is fully consistent with current observational data, there is no observational evidence yet that evolving dark energy is favored over a cosmological constant. This conclusion aligns with recent studies \cite{ye2024bridge}, indicating that thawing quintessence models tend to occupy observationally marginal regions of CPL parameter space.

Looking ahead, upcoming cosmological surveys such as DESI \cite{goldstein2023beyond}, Euclid \cite{mellier2024euclid}, and SKA \cite{xiao2022forecasts} are poised to probe deviations from $\Lambda$CDM at the $10^{-3}$ level or better—precisely the regime where the thawing behavior driven by the loop correction parameter $\delta$ would become detectable. In our framework, the absence of detectable deviation in current data is a natural consequence of the small loop amplitude $|\delta| \ll 1$, which keeps the scalar field effectively frozen over the redshift range $z \in [0, 1]$, yielding $w_\varphi(z) \approx -1$. The resulting observational degeneracy with $\Lambda$CDM highlights the predictive nature of the model: small deviations are not excluded but lie just below present detection thresholds. Moreover, the combined effect of the hilltop potential geometry and logarithmic loop correction induces attractor-like thawing dynamics, wherein a broad range of initial field values converges rapidly onto a common evolutionary path. This renders the model observationally testable in future datasets and distinguishes it from both classical quintessence and Palatini-type constructions, where scalar dynamics are typically either fine-tuned or suppressed at late times. Future constraints on the $(w_0, w_a)$ plane at sub-percent levels—potentially reaching sensitivities of $\mathcal{O}(10^{-5}–10^{-4})$ \cite{akrami2021quintessential}—will directly test the loop-induced deviations predicted by our framework, offering a parameter space where deviations from \(\Lambda\)-CDM would manifest.

However, several limitations of the present framework must be acknowledged. Our analysis remains classical and effective in nature; it does not address the ultraviolet (UV) origin of the $f(R)$ action, nor does it model the post-inflationary reheating process in detail. The late-time vacuum energy scale still requires a small parameter in the Lagrangian, implying that fine-tuning—while alleviated—cannot be entirely avoided, particularly given that the loop amplitude $\delta$ remains observationally unconstrained. Future work could investigate the quantum stability of this construction, its possible embedding within a more fundamental UV-complete theory, and its detailed dynamics after inflation, including particle production and thermalization. Nevertheless, we believe that this model serves as a concrete step forward in the development of unified $f(R)$ cosmologies, unifying inflation and dark energy through a single scalar degree of freedom. 

\section{Declaration of competing interest}
\begin{enumerate}
    \item Author Contributions: All authors contributed equally to the preparation of the manuscript.
    \item Funding: Seed money scheme, Sanction No. CU-ORS-SM-24/29.
    \item Data Availability Statement: Not applicable.
    \item Conflicts of Interest: The authors declare no conflict of interest.
\end{enumerate}

\section{Declaration of competing interest}
\begin{enumerate}
    \item Author Contributions: Pradosh Keshav M.V. conceived the core idea, developed the theoretical framework, and wrote the manuscript. Kenath Arun contributed to the theoretical analysis, numerical implementation, and interpretation of the physical results.
    \item Funding: Seed money scheme, Sanction No. CU-ORS-SM-24/29.
    \item Data Availability Statement: Not applicable.
    \item Conflicts of Interest: The authors declare no conflict of interest.
\end{enumerate}
\section{Declaration of Generative AI and AI-assisted technologies in the writing process}
During the preparation of this work, the author, Pradosh Keshav, used Grammarly to correct grammar mistakes and make some paragraphs more structured. After using this tool/service, the author reviewed and edited the content as needed and take(s) full responsibility for the content of the publication.


\printbibliography

\appendix
\section{Reconstruction and Viability of $f(R)$ from the Jordan Frame Potential}
\renewcommand{\theequation}{A\arabic{equation}}

We provide here a constructive derivation of the gravitational Lagrangian \(f(R)\) corresponding to the scalar potential given in Eq.~\eqref{eq:quintpot}, with the aim of demonstrating how the odd-power structure in Eq.~\eqref{eq:fR_ansatz} emerges from first principles, and how the low-curvature conditions \(f(R_0) = R_0 - 2\Lambda_{\text{eff}}\) and \(f'(R_0) = 1\) are satisfied. We also verify that the model passes solar-system consistency tests by evaluating the scalaron mass and the condition \(0 < R f''(R)/f'(R) < 1\) in the high-curvature regime. Our analysis is carried out for the representative case \(q = 2\) and \(n = 1\), with \(\phi_c = 1\) for clarity. 

We consider the hilltop potential given in Eq.~\eqref{eq:quintpot},
\begin{equation}
V(\phi) = A \left( \left( \frac{\phi}{\phi_c} \right)^2 - 1 \right)^q,
\end{equation}
and specialize to the case where \(q = 2\), \(\phi_c = 1\), yielding
\begin{equation}
V(\phi) = A (\phi^2 - 1)^2, \quad \text{and} \quad V'(\phi) = 4A \phi (\phi^2 - 1).
\end{equation}
We expand near the minimum at \(\phi = 1\) by introducing a small displacement \(\Delta\), such that \(\phi = 1 + \Delta\), with \(|\Delta| \ll 1\). This yields
\begin{align}
V'(1 + \Delta) &= 4A (1 + \Delta) \left[ (1 + \Delta)^2 - 1 \right] \nonumber \\
&= 4A (1 + \Delta)(2\Delta + \Delta^2) \nonumber \\
&\approx 8A \Delta + \mathcal{O}(\Delta^2).
\end{align}
Identifying \(R = V'(\phi)\), we obtain to leading order:
\begin{equation}
R = 8A (\phi - 1) \quad \Rightarrow \quad \phi(R) = 1 + \frac{R - R_0}{8A},
\end{equation}
where we have used the condition \(\phi(R_0) = 1\), corresponding to the minimum of the potential.

We now reconstruct \(f(R)\) by substituting \(\phi(R)\) into the Legendre transform:
\begin{equation}
f(R) = \phi(R)\, R - V\left( \phi(R) \right).
\end{equation}
Here, we define a dimensionless expansion variable \(x \equiv \frac{R - R_0}{8A}\), so that \(\phi(R) = 1 + x\). Then the potential becomes:
\begin{align}
V(\phi(R)) &= A \left[ (1 + x)^2 - 1 \right]^2 = A (2x + x^2)^2 \nonumber \\
&= A (4x^2 + 4x^3 + x^4).
\end{align}
Consequently, the Lagrangian becomes:
\begin{equation}
f(R) = (1 + x)\, R - A(4x^2 + 4x^3 + x^4).
\end{equation}
Expanding this near \(R = R_0\), we find:
\begin{equation}
f(R) = R - \frac{(R - R_0)^2}{8A} + \mathcal{O}\left( (R - R_0)^3 \right).
\end{equation}
This series manifestly satisfies
\begin{equation}
f'(R_0) = 1, \quad f(R_0) = R_0 - 2\Lambda_{\text{eff}},
\end{equation}
with \(\Lambda_{\text{eff}} = \tfrac{1}{2} V(1) = 0\) in this specific case (since \(V(1) = 0\)). However, in the general case with \(q > 2\), \(V(1)\) is nonzero and provides an effective cosmological constant. Higher-order odd powers in \((R - R_0)\) arise upon including additional terms in the expansion of the potential around \(\phi = 1\), which are in turn manifested in higher-order terms in the expansion of \(f(R)\).

\begin{figure*}
    \centering
    \includegraphics[width=0.7\linewidth]{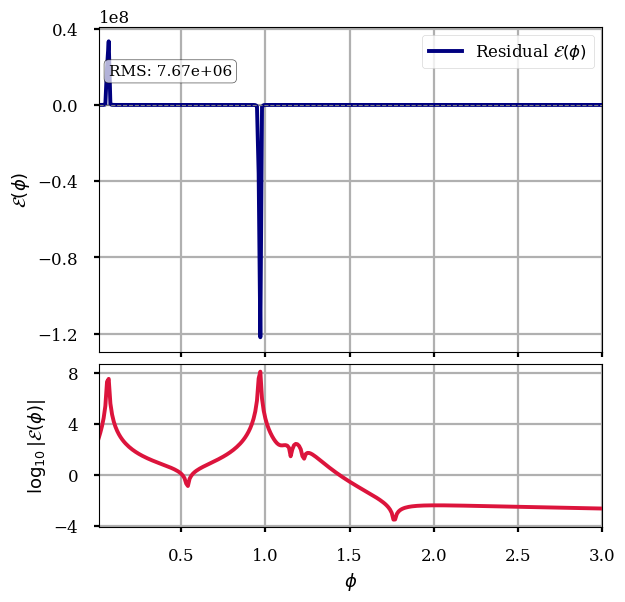}
    \caption{\small  Residual $\mathcal{E}(\phi)$ of the master reconstruction equation, defined as the left-hand side of Eq. \eqref{eq:master_recon}, quantifies the deviation from near-consistency between the Jordan-frame form of $P(\phi)$ in Eq. \eqref{eq:Pphi_def}. A vanishing $\mathcal{E}(\phi)$ implies that the functional form of $P(\phi)$, derived from the odd-power $f(R)$ potential, reproduces the desired cosmic expansion. The observed spikes arise from rapid variations in higher-order derivatives of $P(\phi)$, and the inflection point at $\phi = \phi_c$. Outside these regions, $\mathcal{E}(\phi) \ll 1$, indicating that the reconstructed $f(R)$ form remains consistent with the background dynamics.}
    \label{fig:residualplot}
\end{figure*}

The emergence of higher-order odd powers in \((R - R_0)\) is therefore justified, and motivates the rational ansatz of Eq.~\eqref{eq:fR_ansatz}, which we repeat here for clarity:
\begin{equation}
f(R) = -\frac{(R - R_0)^{2n+1} + R_0^{2n+1}}{f_0 + f_1 \left[ (R - R_0)^{2n+1} + R_0^{2n+1} \right]}.
\end{equation}
For \(n = 1\), its Taylor expansion about \(R = R_0\) yields:
\begin{equation}
f(R) = R_0 - 2\Lambda_{\text{eff}} + (R - R_0) + \mathcal{O}\left((R - R_0)^3\right),
\end{equation}
recovering the correct GR limit. To confirm that this rational form of $f(R)$ also yields consistent background dynamics, we evaluate the residual $\mathcal{E}(\phi)$ of the master reconstruction equation using the associated Jordan-frame function $P(\phi)$ from Eq. \eqref{eq:Pphi_def}. As shown in Fig. \ref{fig:residualplot}, the residual remains small across most of the field range, validating that the odd-power structure embedded in the ansatz reproduces the desired expansion history. 

We now verify that this form is consistent with solar-system tests. The viability condition given by Odinstov et al \cite{odintsov2023recent}:
\begin{equation}
0 < \frac{R f''(R)}{f'(R)} < 1 \label{eq:viability}
\end{equation}
must hold in the high-curvature regime to avoid fifth-force constraints. We define:
\begin{align}
N(R) &= (R - R_0)^{2n+1} + R_0^{2n+1}, \\
D(R) &= f_0 + f_1 N(R),
\end{align}
so that
\begin{equation}
f(R) = -\frac{N(R)}{D(R)}, \qquad f'(R) = -\frac{f_0 N'(R)}{D(R)^2},
\end{equation}
with
\begin{align}
N'(R)  &= (2n + 1)(R - R_0)^{2n}, \\
N''(R) &= (2n + 1)(2n)(R - R_0)^{2n - 1}.
\end{align}
Then,
\begin{equation}
f''(R) = -f_0 \left[ \frac{N''(R)}{D^2} - \frac{2 f_1 N'(R)^2}{D^3} \right].
\end{equation}
Evaluating at \(R = R_0\), we find:
\begin{align}
N'(R_0) &= 0, & N''(R_0) &= 0 \nonumber \\
\Rightarrow \quad f'(R_0) &= 1, & f''(R_0) &= 0.
\end{align}
Thus,
\begin{equation}
\lim_{R \to R_0} \frac{R f''(R)}{f'(R)} = 0,
\end{equation}
satisfying the condition in Eq. \eqref{eq:viability}. Moreover, the scalaron mass \(m^2(R) = \frac{f'(R)}{3f''(R)}\) diverges as \(f''(R_0) \to 0^+\), indicating that the scalar degree of freedom becomes infinitely heavy in high-curvature regimes. This ensures the suppression of any fifth-force effects and confirms consistency with solar-system constraints such as the Cassini bound.

\section{ One-Loop Effective Potential and the Logarithmic Correction}
\renewcommand{\theequation}{B\arabic{equation}}

In this appendix, we derive the logarithmic correction to the scalar potential arising at one-loop order, which motivates the form of the loop-corrected potential used in Eq.~\eqref{eq:upotenfinal}. The correction originates from integrating out quantum fluctuations of a light scalar field in a curved spacetime background, in analogy with the Coleman–Weinberg mechanism in flat space and its covariant extension to gravitational systems. In the Einstein frame, the scalar potential \( U(\varphi) \) receives radiative corrections from its own quantum fluctuations, or those of other fields coupled to it. 

The one-loop effective potential in flat spacetime is given by the standard expression \cite{coleman1973radiative}:
\begin{equation}
U_{\text{1-loop}}(\varphi) = U_0(\varphi) + \frac{1}{64\pi^2} M^4(\varphi) \ln\left( \frac{M^2(\varphi)}{\mu^2} \right),
\end{equation} where lower-order constants and scheme dependence are absorbed into \( U_0(\varphi) \) (which is the classical potential) and the renormalization scale \(\mu\). In our case, the effective field-dependent mass squared \(M^2(\varphi)\) is identified with the second derivative of the Einstein-frame potential with respect to \(\varphi\), i.e.,
\begin{equation}
M^2(\varphi) = \frac{d^2 U_0}{d\varphi^2},
\end{equation}
where \(U_0(\varphi)\) arises from the conformal transformation of the Jordan-frame action, given by
\begin{equation}
U_0(\varphi) = \frac{V(\varphi)}{2\kappa^2 \varphi^2},
\end{equation}
as in Eq.~\eqref{eq:15} of the main text. For the hilltop form \( V(\phi) = A \left( \left(\frac{\phi}{\phi_c} \right)^2 - 1 \right)^q \), this yields
\begin{equation}
U_0(\varphi) = \frac{A}{2\kappa^2 \varphi^2} \left[ \left( \frac{\varphi}{\varphi_c} \right)^2 - 1 \right]^q.
\end{equation}

The loop-corrected potential is then
\begin{equation}
U(\varphi) = U_0(\varphi) + \frac{1}{64\pi^2} \left( \frac{d^2 U_0}{d\varphi^2} \right)^2 \ln\left( \frac{1}{\mu^2} \frac{d^2 U_0}{d\varphi^2} \right).
\end{equation}
Since this form is highly model-dependent and difficult to treat analytically for arbitrary \(q\), we adopt a phenomenologically motivated simplification by assuming that the dominant radiative correction arises from an effective log-type deformation of the potential. Accordingly, we write the effective potential as
\begin{equation}
U(\varphi) = U_0(\varphi) + \mathcal{Z} \ln\left( \frac{\varphi}{\varphi_0} \right),
\end{equation}
where \(\varphi_0\) is a normalization scale (absorbing the renormalization scale \(\mu\)), and \(\mathcal{Z} < 0\) is a small dimensionful coefficient that controls the strength of the correction. This form ensures stability of the potential at late times and mimics the loop-induced flattening observed in radiative corrections to scalar potentials. A similar form arises in effective gravity theories with light quantum fields propagating on a curved background, as discussed in Refs.~\cite{inagaki2005one, nojiri2000thermodynamics, elizalde1994renormalization, elizalde2007quantum, elizalde1996effective, elizalde1998one}.

\section{Einstein-frame Potential from Loop-Corrected \( f(R) \) Gravity}
\renewcommand{\theequation}{C\arabic{equation}}
We outline the derivation of the Einstein-frame scalar potential arising from Eq. \eqref{eq:fRloop} in the main text. This form of \( f(R) \) arises in the effective gravitational action from quantum corrections due to vacuum polarization, e.g., via trace anomalies in curved spacetime \cite{mottola2010trace, mottola2011new}. We proceed to map this theory to its scalar-tensor representation via a Legendre transformation and then conformally transform to the Einstein frame.

To introduce a scalar degree of freedom, we map this theory to a scalar-tensor representation via a Legendre transformation. Define the auxiliary field \( \phi \equiv f'(R) \) as:
\begin{equation}
\phi \equiv f'(R) = 1 + 2\alpha R + \beta\left[2R \ln\left( \frac{R}{\mu^2} \right) + R\right].
\label{eq:scalarphi}
\end{equation}
This relation is transcendental and cannot be inverted analytically in closed form. However, for large curvature \( R \gg \mu^2 \), corresponding to large \( \phi \), the logarithmic term dominates, and we approximate:
\begin{equation}
\phi(R) \approx 2\beta R \ln\left( \frac{R}{\mu^2} \right),
\end{equation} where solving and inverting iteratively gives \( R(\phi) \sim \phi/\left(2\beta \ln \phi\right) \) up to subleading corrections. Although the inversion \( R(\phi) \) is not analytically tractable in closed form, it suffices to invert this relation approximately in the asymptotic regime, which is valid for large \( \phi \) or equivalently large \( \varphi \) in the Einstein frame.

The conformal transformation to the Einstein frame is implemented via \( \tilde{g}_{\mu\nu} = \phi \, g_{\mu\nu} \), and the canonically normalized scalar field is given by
\begin{equation}
\varphi = \sqrt{\frac{3}{2}} M_{\rm Pl} \ln \phi,
\quad \Rightarrow \quad \phi = e^{\sqrt{2/3}\, \varphi / M_{\rm Pl}}.
\label{eq:phidef}
\end{equation}
The Einstein-frame potential is then obtained through
\begin{equation}
V(\varphi) = \frac{R \phi - f(R)}{2\kappa^2 \phi^2},
\label{eq:EinsteinPotentialGeneral}
\end{equation}
where all quantities on the right-hand side are understood as functions of \( R \), and ultimately of \( \phi \) and \( \varphi \). Substituting Eqs. \eqref{eq:fRloop} and \eqref{eq:scalarphi}, we compute:
\begin{align}
V(\phi) &= \frac{1}{2\kappa^2 \phi^2} \left[ R \phi - \left( R + \alpha R^2 + \beta R^2 \ln\left( \frac{R}{\mu^2} \right) \right) \right] \nonumber \\
&= \frac{1}{2\kappa^2 \phi^2} \left[ R (\phi - 1) - R^2 \left( \alpha + \beta \ln\left( \frac{R}{\mu^2} \right) \right) \right].
\label{eq:Vphiint}
\end{align}

Using the asymptotic form of \( R \sim \phi / (2\beta \ln(\phi/\mu^2)) \) and replacing \( \phi \) via Eq.~\eqref{eq:phidef}, the potential simplifies to a convenient form:
\begin{dmath}
V(\phi) \sim \frac{1}{2\kappa^2} \frac{R\phi - f(R)}{\phi^2}
\sim \frac{1}{2\kappa^2} \left[ \frac{1}{\phi^2} R \cdot \phi - \frac{R^2}{\phi^2} \left( \alpha + \beta \ln\phi \right) \right].
\end{dmath}
This gives, up to an overall normalization,
\begin{equation}
V(\varphi) \propto \phi^{-2} \left[ \phi \cdot R - R^2 \left( \alpha + \beta \ln \phi \right) \right].
\label{eq:Vphistart}
\end{equation}
Now, using Eq. \eqref{eq:phidef}, one has \( \ln \phi = \sqrt{2/3} \, \varphi/M_{\rm Pl} \), and \( \phi^{-2} = e^{-2\sqrt{2/3} \, \varphi/M_{\rm Pl}} \). Substituting these expressions into Eq.~\eqref{eq:Vphistart}, and absorbing the subleading logarithmic dependence of \( R \sim \phi / \ln \phi \), the Einstein-frame potential takes the form:
\begin{equation}
V(\varphi) \sim e^{-2\sqrt{2/3}\, \varphi/M_{\rm Pl}} \left[ 1 + \delta \ln\left( \frac{\varphi}{\varphi_0} \right) \right],
\label{eq:Vfinalform}
\end{equation}
where \( \delta \sim \beta/\alpha \ll 1 \) is the relative strength of the one-loop correction and \( \varphi_0 \) is a renormalization scale. This form arises naturally by noting that \( \ln \phi \sim \varphi \), and hence any logarithmic deviation in \( f(R) \) results in a subleading \( \varphi \)-dependent modulation to the pure exponential potential. Restoring normalization, the scalar potential in the Einstein frame, including one-loop effects, takes the form:
\begin{equation}
V(\varphi) = V_0\, \exp\left[ -\mu \varphi \right] \left[ 1 + \delta \ln\left( \frac{\varphi}{\varphi_0} \right) \right],
\quad \mu = \frac{2\sqrt{2/3}}{M_{\rm Pl}},
\end{equation}
which generalizes the tree-level Starobinsky potential with logarithmic flattening induced by quantum corrections.
\end{document}